\documentclass[sigconf]{acmart} 
\AtBeginDocument{
  }

\copyrightyear{2026}
\acmYear{2026}
\setcopyright{cc}
\setcctype{by}
\acmConference[CHI '26]{Proceedings of the 2026 CHI Conference on Human Factors in Computing Systems}{April 13--17, 2026}{Barcelona, Spain}
\acmBooktitle{Proceedings of the 2026 CHI Conference on Human Factors in Computing Systems (CHI '26), April 13--17, 2026, Barcelona, Spain}
\acmPrice{}
\acmDOI{10.1145/3772318.3790987}
\acmISBN{979-8-4007-2278-3/2026/04}

\usepackage[font=bf]{caption}
\usepackage[english]{babel}
\usepackage{csquotes}
\usepackage{float}

\AtBeginDocument{
  }

\begin{document}

\author{Qing Hu}
\affiliation{
  \department{Human-Computer Interaction Institute} 
  \institution{Carnegie Mellon University}
  \city{Pittsburgh}
  \state{PA}
  \country{USA}
}
\email{dianehu@andrew.cmu.edu}

\author{Qing Xiao}
\affiliation{
 \department{Human-Computer Interaction Institute} 
  \institution{Carnegie Mellon University}
  \city{Pittsburgh}
  \state{PA}
  \country{USA}
}
\email{qingx@cs.cmu.edu}

\author{Hancheng Cao}
\affiliation{
  \department{Goizueta Business School} 
  \institution{Emory University}
  \city{Atlanta}
  \state{GA}
  \country{USA}
}
\email{hancheng.cao@emory.edu}

\author{Hong Shen}
\affiliation{
  \department{Human-Computer Interaction Institute} 
  \institution{Carnegie Mellon University}
  \city{Pittsburgh}
  \state{PA}
  \country{USA}
}
\email{hongs@cs.cmu.edu}

\title[Manager Clone Agents in the Future Workplace]{When Your Boss Is an AI Bot: Exploring Opportunities and Risks of Manager Clone Agents in the Future Workplace}

\begin{teaserfigure}
  \centering
  \includegraphics[width=0.9\textwidth]{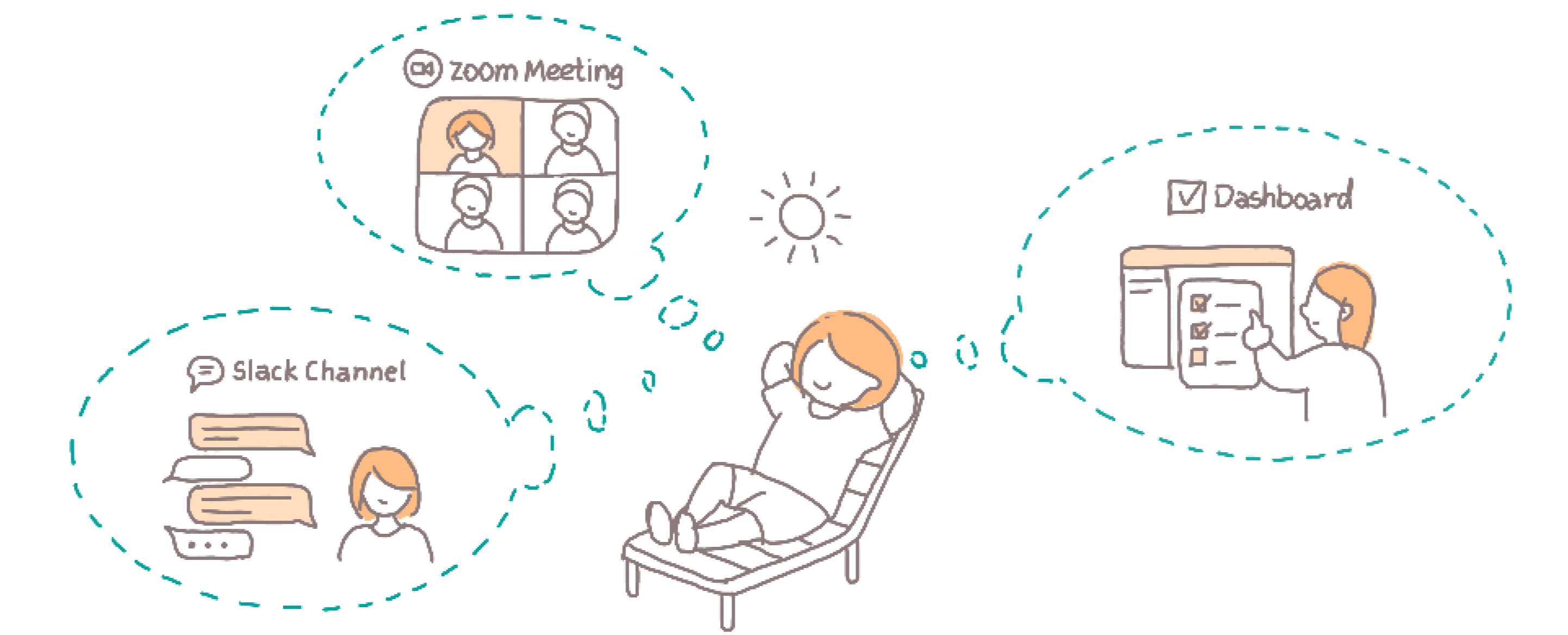}
  \caption{Manager Clone Agents that replicate a manager’s symbolic, social, and communicative presence. They act as digital clones, mimicking a manager’s appearance, behaviors, and communication style, and can also serve into functional delegates that perceive context, reason through tasks, and execute complex actions without intervention such as attending meetings, facilitating discussion, and making decisions.}
  \Description{The illustration with orange accents shows a human manager reclining on a lounge chair under the sunshine at the bottom center. Three large bubbles around them portray the Manager Clone Agent working in parallel: the upper-left bubble depicts a 2×2 video-call grid with one Manager Clone Agent among worker tiles; the lower-left bubble shows a Slack-like chat where workers send messages and the agent replies; the right bubble shows a task checklist/dashboard where an agent avatar checks items off. Small bubbles coming out from each major bubble indicate delegation.}
  \label{fig:mca}
\end{teaserfigure}

\begin{abstract}
As Generative AI (GenAI) becomes increasingly embedded in the workplace, managers are beginning to create Manager Clone Agents—AI-powered digital surrogates trained on their work communications and decision patterns to perform managerial tasks on their behalf. To investigate this emerging phenomenon, we conducted six design fiction workshops (n = 23) with managers and workers, in which participants co-created speculative scenarios and discussed how Manager Clone Agents might transform collaborative work. We identified four potential roles that participants envisioned for Manager Clone Agents: proxy presence, informational conveyor, productivity engine, and leadership amplifier, while highlighting concerns spanning individual, interpersonal, and organizational levels. We provide design recommendations envisioned by both parties for integrating Manager Clone Agents responsibly into the future workplace, emphasizing the need to prioritize workers’ perspectives and nurture interpersonal bonds while also anticipating alternative futures that may disrupt managerial hierarchies.
\end{abstract}



\maketitle

\section{Introduction}

Recent advances in AI have given rise to the phenomenon of managers “cloning themselves” to delegate routine and strategic work such as attending meetings, exchanging information with employees and stakeholders, and assisting in decisions~\cite{forsdick2025aiavatars,raconteur2023ai,Gani2025ai_hotline}. Unlike algorithmic systems that primarily optimize decisions or workflows~\cite{lee2015working}, \textit{Manager Clone Agents} replicate a manager’s symbolic, social, and communicative presence. They act as digital clones~\cite{Liang2025AIClones,lee_speculating_2023}, mimicking a manager’s appearance, behaviors, and communication style, and can also serve into functional delegates~\cite{cheng_conversational_2025,journey2025aiAgentsExecutiveProductivity} that perceive context, reason through tasks, and execute complex actions without intervention (see~\autoref{fig:mca}).

In practice, such systems are already emerging in heterogeneous forms, from virtual avatars that project an executive’s presence to lightweight assistants embedded in workflows that handle approvals or reports. These designs differ in fidelity, autonomy, but collectively signal a rapidly expanding design space. High-profile executives such as Nvidia CEO Jensen Huang~\cite{nvidia2022omniverse} and LinkedIn co-founder Reid Hoffman~\cite{heygen-reidai} have released their AI avatars in public, while companies like Zoom~\cite{peters2025techceos} and Otter.ai~\cite{boyle2025aiCEO} pilot digital stand-ins for meetings and briefings. Surveys further suggest that adoption is cascading beyond the C-suite into middle management~\cite{maslworld2025middlemanager,Langley2025Lattice}, aided by falling implementation costs and expectations of measurable productivity gains.

Yet, alongside this momentum lie pressing concerns. Because managers shape decision-making, resource distribution, and workplace culture~\cite{Mintzberg_1973}, delegating aspects of their authority, whether symbolically or functionally, to AI agents raises difficult questions of trust, fairness, and accountability. Workers worry about displacement and the erosion of authentic communication~\cite{Bonos_Abril_2025}. These tensions make it urgent to study how such agents alter leadership, collaboration, and organizational dynamics, and to identify pathways for their responsible design and adoption.

Existing HCI studies have begun to examine how AI Clones (personalized AI mimics that approximate individuals given limited training data~\cite{Liang2025AIClones}) coexist with their human originals, often focusing on questions of identity, self-perception and other interpersonal relationships~\cite{hwang_whose_2024,huang_mirror_2025,wang_one_2025,lee_speculating_2023}. 
Building on this work, we use the term \textit{Manager Clone Agents} to refer broadly to a family of emerging systems, ranging from symbolic clones that replicate presence to functional delegates that automate managerial tasks. Positioned at the intersection of authority and collaboration, they combine task-oriented delegation with the symbolic and communicative authority of human managers. This dual capacity makes them distinct from prior managerial technologies~\cite{lee2015working,wolf2025one,Sowa_Przegalinska_Ciechanowski_2020}, but we know little about how their human-like qualities will reshape teamwork within organizations.

Therefore, in this paper, we ask:
\begin{itemize}
    \item \textbf{RQ1:} What opportunities and supportive roles do managers and workers envision for Manager Clone Agents in the future workplace?  
    \item \textbf{RQ2:} What challenges and risks do managers and workers associate with the use of Manager Clone Agents in the future workplace?  
    \item \textbf{RQ3:} How do managers and workers envision the future design of Manager Clone Agents to address these opportunities and risks?  
\end{itemize}

To explore these questions, we conducted a series of design fiction workshops \cite{lyckvi2018role,muller2020understanding,ma2025speculative} with managers and workers, asking participants to imagine organizational futures where Manager Clone Agents were widely adopted. Design fiction, a method widely used in HCI, was chosen because Manager Clone Agents remain speculative technologies at this moment for most managers and workers: it allows participants to engage with fictional scenarios to articulate hopes, conflicts, and concerns without being constrained by today’s technical limits. In this paper, we define “\textit{managers}” as organizational leaders at multiple levels, including executives, directors, and middle managers, and “\textit{workers}” as their employees or direct reports who experience the effects of managerial authority. In the workshops, both groups created scenarios of Manager Clone Agents use and then co-reflected on the opportunities, tensions, and uncertainties these agents might generate in organizational life. 

Our findings reveal a fundamental tension between the promise and risks of Manager Clone Agents. On the one hand, participants envisioned supportive roles, acting as proxy presences in meetings, conveying information across organizational layers, automating routine tasks, and amplifying managerial guidance (\textbf{\textit{RQ1}}). On the other hand, these same roles raised risks across multiple levels: at the individual level, from different positions both managers and workers experienced shared anxieties about accountability and irreplaceability; at the interpersonal level, both sides pointed to fragile trust, blurred authenticity, and reduced informal contact; and at the organizational level, participants worried that efficiency gains could flatten hierarchies and weaken belonging (\textbf{\textit{RQ2}}). Participants, therefore, stressed that responsible design must foreground workers’ voices, preserve communication and social cohesion, and allow trust to develop gradually through clear boundaries on the capabilities of manager clone agents, while remaining attentive to speculative alternative futures that can affect managerial authority (\textbf{\textit{RQ3}}).

This study makes three key contributions to the HCI and CSCW community:
\begin{itemize}
    \item First, we present an empirical vision of potential roles that emerging Manager Clone Agents could undertake. Drawing on perspectives from both managers and workers, we surface the opportunities and risks these roles create across individual, interpersonal, and organizational levels in future workplaces.  
    \item Second, we distinguish Manager Clone Agents from prior algorithmic management systems by showing how their personalization and symbolic authority recast debates on authenticity and automation, while reframing both managerial work and worker-manager relations as relational and organizational challenges of AI automation.
    \item Finally, we surface design directions for developing Manager Clone Agents that build trust and support collaborations within diverse team and organizational contexts, including those challenging the current hierarchies.  
\end{itemize}

\section{Background: Emerging Manager Clone Agents in Workplaces} 

Recent advances in artificial intelligence have enabled the rise of Manager Clone Agents. Executives are beginning to use these tools to manage heavy schedules, from financial briefings to routine worker inquiries. As early as 2021, Nvidia CEO Jensen Huang released an AI avatar powered by Omniverse Avatar~\cite{nvidia2022omniverse}. Since then, startups such as Personal AI\footnote{https://www.personal.ai/}
, Delphi\footnote{https://www.delphi.ai/}
, and Tavus\footnote{https://www.tavus.io/}
 have built platforms that train avatars on leaders’ speeches and meetings to create chatbots, voice agents, and video avatars. Investment has followed quickly: Delphi raised \$16M in June 2025~\cite{delphi2025seriesA}, HeyGen \$60M~\cite{heygen2025seriesA}, and ElevenLabs \$180M at a \$3.3B valuation~\cite{elevenlabs2025seriesC}, while Synthesia reached \$2.1B~\cite{synthesia2025seriesD}. High-profile leaders such as Linkedin co-founder Reid Hoffman, Zoom CEO Eric Yuan, and Khosla Ventures managing director Keith Rabois have also begun experimenting with these systems~\cite{heygen-reidai,patel_zoom_ceo_2024,morrone2025ai}, using them for financial updates~\cite{bort2025klarna,peters2025techceos}, meeting attendance~\cite{boyle2025aiCEO}, or routine staff communication~\cite{personalai_salomon_2025,sequoia2025delphi,Langley2025Lattice,cuthbertson2023company}. The adoption of AI agents in workplaces is also accelerating. A PwC survey of 308 U.S. managers in April 2025 reported that 79\% of companies already use AI agents, and 88\% plan to expand budgets within a year~\cite{pwcAIAgentSurvey2025}. Falling costs are lowering entry barriers~\cite{FiferFeagans2025}, positioning Manager Clone Agents to spread from executives to middle and line managers in the near future~\cite{digitaldefynd2025ai}. Yet concerns remain. Because managerial roles shape decisions, resources, and culture, delegating them to AI raises questions of trust, accountability, and fairness~\cite{Sowa_Przegalinska_Ciechanowski_2020}. Employees expressed concerns about the erosion of authentic communication, job loss, and the widening gap between executive privilege and labor precarity~\cite{writer2025generativeai, morrone2025ai}. These tensions highlight the need to better understand how Manager Clone Agents affect the future workplace.

\section{Related Work}
First, we review the scholarship on managerial roles in the workplace, which highlights the complex mix of decision-making, coordination, and relational care that define managerial labor (see \autoref{lr1}). Second, we discuss HCI and CSCW work that examines the role of technologies in supporting managerial work, from algorithmic management to collaborative robots and AIs (see \autoref{lr2}). Third, we examine the early work on Clone Agents, focusing on their potential to extend human presence while raising new concerns about identity and authenticity (see \autoref{lr3}). 

\subsection{Managerial Roles in the Workplace} \label{lr1}

Managers occupy a distinctive position in organizational life. Unlike most employees, they are responsible for abstract decision-making, conflict coordination, and emotional motivation, which makes their tasks qualitatively different from those of their subordinates~\cite{dler2021importance}. Classic organizational theory frames managerial work as a constellation of intertwined roles: interpersonal (figurehead, leader, liaison), informational (monitor, disseminator, spokesperson) and decisional (entrepreneur, disturbance handler, resource allocator, negotiator)~\cite{Mintzberg_1973}. Yet these responsibilities accumulate into heavy burdens. Managers already face fragmented meetings, emails, and constant communication that limit time for reflection and strategy~\cite{stray2020understanding,dler2021importance}. Recent shifts in digital technologies that multiply channels and complicate coordination~\cite{Andersone_Nardelli_Ipsen_Edwards_2022}, along with the demands of remote and hybrid work \cite{morrison2020challenges,choudhury2023virtual} have further amplified these strains.

Beyond the structural demands of the role, effective management depends on the individual ways managers enact their responsibilities. Leadership research highlights how different decision-making styles (e.g., autocratic, democratic) produce divergent effects on team performance, satisfaction, and trust~\cite{YuklVanFleet1992}. Some managers also construct personal narratives that align organizational goals, thereby legitimizing their authority~\cite{rostron2022hero}. These perspectives underscore that effectiveness in management hinges less on a fixed formula and more on how managers’ personal identities and the relational dynamics they cultivate. Organizational efforts to ease workload, such as appointing chiefs of staff, executive assistants, or project management offices, reinforce this point: the success of such delegation hinges on trust, specifically on whether decision-making comes from trusted internal leaders or less familiar external consultants~\cite{meagher2020worker,tamkin}. 

This body of work shows that managers must simultaneously juggle strategic, operational and interpersonal duties, while also enacting identities that sustain their authority. The sophistication and personalization of management, and the role it plays in team dynamic, call for more nuanced and effective support. In this paper, we explore Manager Clone Agents as a novel possibility: digital delegate trained to mimic a manager’s communication style, decision-making patterns, and even personal presence. 

\subsection{Technologies that Support Managerial Roles} \label{lr2}
Scholars in HCI and CSCW have long examined how algorithmic systems, robots and other forms of AIs assume partial managerial roles, reshaping workflows and authority structures. One prominent line of research is algorithmic management~\cite{lee2015working,lee_algorithmic_2017,lee_procedural_2019,green2019principles,cao2021my}, where decision-making roles are automated through software and real-time data analytics. It has underscored how algorithmic oversight reconfigures worker-manager relations, offering both tighter control and diminished autonomy for human supervisors. Another related stream focuses on how collaborative robots (cobots)~\cite{cheon2022working} reshaped roles and trust, and that an ideal image of a robot leader needs to be tailored to individuals and set appropriate expectation management~\cite{wolf2025one}. 

AI is no longer merely an assistive tool that teams use to accomplish tasks; it is increasingly becoming an active part of the team, functioning as an agent that works alongside people in their everyday collaboration~\cite{ofem_tool_2025,dennis_ai_2023}. At the executive level, recent scholars have begun theorizing about the Chief AI Officer (CAIO) role as a strategic necessity~\cite{Schmitt_2024}. Research in Human–AI Teaming (HAT) further shows that when AIs join teams, the initial expectation they received in terms of communication capabilities, human-like behaviors and performance shape how trust and power are formed and developed among the team~\cite{duan2024understanding,zhang2021ideal,flathmann2024empirically}. Building on this, Sadeghian et al. highlights how the interaction patterns between workers and AI systems directly influence workers’ sense of satisfaction and the meaning they derive from their roles~\cite{sadeghian2024soul}. These studies establish that technologies, whether algorithms, collaborative robots, or AI leadership frameworks, do more than optimize processes: they reshape authority, identity, and social dynamics within organizations.

At the same time, the introduction of collaborative and managerial technologies provokes persistent concerns among workers about automation, control, and trust. Grudin’s well-known observation in 1988 has already warned that groupware adoption often fails when those responsible for entering data do not benefit directly from its use, or when organizational cultures prioritize competition over sharing~\cite{grudin1988cscw}. Recent case studies extend these concerns to AI-based systems, showing that successful adoption depends on managers addressing both cognitive trust (clear understanding of how AI works and what its limitations are)~\cite{ofem_tool_2025} and emotional trust (a sense of safety, fairness, and empathy in its use)~\cite{vuori2025s,lee2018understanding,rosenblat2016algorithmic}. In particular, the negative impacts of algorithmic management often arise from systems that favor managerial efficiency over an inclusive design that centers on worker well-being~\cite{zhang2022a,spektor2023charting}. In many cases, the success or failure of management technologies hinges not only on technical accuracy but also on how well they align with cultural values of fairness, agency, and inclusiveness.

These streams of research reveal that technological interventions in management often boost efficiency and control but risk rigidity, impersonality, and worker resistance. They also prioritize procedural and physical aspects over relational and cultural dynamics. As collaborative technologies develop, leadership requires flexibility, trust-building, and shared meaning. Manager Clone Agents, unlike rigid algorithms, offer personalized support by simulating real managers’ styles to mediate tasks, communication and collaboration.

\subsection{Clone Agents in Workplace} \label{lr3}
With the rise of GenAI, AI agents have emerged as a distinct category of intelligent digital assistants, capable of autonomously performing tasks on behalf of users and mimicking aspects of human interaction~\cite{park_generative_2023,park2024generative,10.1145/3613904.3642114}. A growing stream of research explores their role in teamwork, where AI agents can support managerial functions ranging from scheduling~\cite{cranshaw2017calendar} to decision-making~\cite{li2025metaagents}. Systems such as Cobot-style workflow assistants~\cite{Sowa_Przegalinska_Ciechanowski_2020} illustrate how agents can reduce leaders’ follow-up burdens. Other studies highlight how conversational agents can facilitate group discussions, such as encourageing equal participation~\cite{do2022should,kim2020bot}, and team cohesion~\cite{benke2020chatbot}. These developments signal the growing managerial relevance of AI agents in collaborative contexts.

Building on this trajectory, Clone Agents represent a special subtype of AI agents. Unlike generic digital assistants, Clone Agents are designed to replicate not only an individual’s decision logic~\cite{truby_human_2021} but also their voice, appearance and conversational style~\cite{dobre_nice_2022,leong_dittos_2024,shamekhi_face_2018}. Scholars have begun to conceptualize these entities as AI clones—digital representations trained on personal data that can interact with others as if they were a real-world individual~\cite{lee_speculating_2023}. Early HCI and CSCW systems hinted at this direction: Time-Travel Proxy enabled asynchronous participation through pre-recorded video~\cite{tang_time_2012}, and Lee \& Takayama showed how mobile remote presence let remote workers “live and work” with colleagues almost as if physically there~\cite{lee_now_2011}. More recent work extends these ideas: Dittos created embodied proxies for real-time meetings~\cite{leong_dittos_2024}, and Cheng et al. studied how delegating conversational autonomy to agents affects multitasking in group interactions~\cite{cheng_conversational_2025}.

Manager Clone Agents, unlike rigid algorithms, offer personalized support by simulating real managers’ styles to mediate tasks, communication and collaboration~\cite{forsdick2025aiavatars}. Whereas digital twins~\cite{vainionpaa_hci_2022,lv_advanced_2023} primarily model physical or organizational assets, algorithmic management~\cite{lee2015working,zhang2022a} emphasizes efficiency and control, and avatars~\cite{gao_effects_2020,weidner_systematic_2023} or cobots~\cite{weiss_cobots_2021,wolf2025one} focus on embodiment and interaction, Clone Agents are distinctive in that they merge personalization with symbolic authority, functioning as proxies that extend a specific manager’s presence and decision-making into organizational life. This dual capacity makes them qualitatively different from prior technologies, raising new questions about authenticity, legitimacy, and relational trust in AI-mediated management.

In terms of identity and authenticity, a core concern is that Clone Agents can exploit or steal an individual’s identity~\cite{huang_mirror_2025,wang_one_2025}. Some fear “\textit{doppelganger-phobia}”: one’s digital clone might replace or outperform them, potentially altering relationships with colleagues, employers, or even loved ones~\cite{lee_speculating_2023}. Trust also becomes vulnerable in environments where clones and deepfakes circulate as synthetic media erodes confidence in visual and auditory evidence~\cite{johnson_what_2021,kaate_how_2023,hwang_whose_2024}. Finally, privacy and consent emerge as pressing challenges, as digital likenesses may be generated or deployed without authorization~\cite{umbach_non-consensual_2024,lee_speculating_2023}. 

Clone Agents hold significant promise as a novel form of AI agent in the future workplace, offering new ways to extend presence, delegate routine managerial work, and maintain continuity in collaborative settings. Yet, their uniqueness also introduces profound risks for accountability, trust, and word culture. Within this landscape, Manager Clone Agents mark a critical turning point, concentrating these opportunities and risks in ways that directly challenge how authority and collaboration are organized in the future workplace.
\section{Method}
To investigate how Manager Clone Agents might reshape collaboration in the workplace, we conducted six two-hour design fiction workshops with 23 participants in total, comprising both managers and workers in each session. Each workshop brought together three to five participants who co-created speculative scenarios and reflected on the opportunities, conflicts, and risks that Manager Clone Agents might generate in organizational contexts. Design fiction was particularly well-suited to this study because Manager Clone Agents remain speculative technologies \cite{muller2020understanding,lyckvi2018role}; by situating participants in fictional, yet plausible scenarios, we were able to elicit relational concerns and organizational imaginaries that might not emerge through surveys or interviews focused only on current practices. 

\subsection{Participants}  

In this study, we define managers as individuals who hold formal authority to supervise, coordinate or make decisions affecting others’ work, and workers as those who carry out tasks under managerial oversight without primary supervisory responsibilities. We specifically recruited both managers and workers to capture perspectives from those exercising managerial authority as well as those subject to it.

We recruited 23 participants through professional networks, workplace mailing lists, and social media outreach. To ensure diversity, we used purposive sampling \cite{campbell2020purposive} to include individuals from a range of industries (e.g., technology, design, education, media, healthcare, manufacturing, and government) and with varied levels of work experience (from less than one year to over a decade). \autoref{tab:participants} summarizes participants’ demographic details.

\begin{table*}[ht]
\centering
\caption{Workshop Participants’ Demographic.}
\label{tab:participants}

\footnotesize\textit{Note:} The “ID” prefix indicates participant’s role (M = Manager, W = Worker). Participants grouped under each section participated in the same Workshop.

\vspace{4pt}

\begin{tabular}{llllll}

\toprule
\textbf{ID} & \textbf{Role \& Sex} & \textbf{Age} & \textbf{Work Exp.} & \textbf{Team (Industry)} & \textbf{Team Size}\\
\midrule
W1  & Worker, Female  & 28 & 1--2 years   & Product Team (Design)                         & 10-15 \\
M2  & Manager, Male   & 25 & 1--2 years   & Risk Management Team (Renewable Energy)      & 20-25 \\
W3  & Worker, Female  & 24 & 1--2 years   & Design Team (Automotive Manufacturing)       & 20-25 \\
M4  & Manager, Female & 29 & 3--5 years   & Design Team (Automotive Manufacturing)       & 10-15 \\
\midrule
M5  & Manager, Male   & 25 & 3--5 years   & Technical Production Team (Television Production) & 30-35 \\
M6  & Manager, Male   & 24 & 1--2 years   & Event Planning Team (Media)                 & 5-10 \\
W7  & Worker, Female  & 24 & 0--1 year    & Data Production Team (Internet) & 10-15 \\
W8  & Worker, Female  & 26 & 0--1 year    & Research Team (Technology)                  & 5-10 \\
\midrule
M9  & Manager, Female & 30 & 1--2 years   & Product Team (Education)                    & 5-10 \\
M10 & Manager, Male   & 28 & 3--5 years   & Cross-Functional Team (Internet)  & 5-10 \\
W11 & Worker, Male    & 25 & 0--1 year    & Research Team (Technology and Education)    & <5 \\
W12 & Worker, Female  & 26 & 0--1 year    & Product Team (Internet)                     & 5-10 \\
M13 & Manager, Female & 27 & 1--2 years   & Project Execution Team (Government Agencies)& <5 \\
\midrule
W14 & Worker, Male    & 28 & 3--5 years   & Editorial Team (Digital Media)              & 5-10 \\
W15 & Worker, Male    & 27 & 3--5 years   & Design Team (Healthcare)                    &  5-10  \\
M16 & Manager, Female & 30 & 3--5 years   & Cross-functional Team (Professional Development) & 10-15 \\
M17 & Manager, Female & 31 & 5--10 years  & Marketing Team (Marketing)                  & 5-10 \\
\midrule
M18 & Manager, Male   & 32 & 5--10 years  & Software QA Team (Foundation Model Infrastructure) & 10-15 \\
M19 & Manager, Male   & 45 & >10 years    & Product Team (Information Technology)       & >50 \\
W20 & Worker, Female  & 25 & 1--2 years   & Product Team (Racing Equipment)             & 5-10 \\
\midrule
M21 & Manager, Male   & 39 & 5--10 years  & Communication Design Team (Design)          & 5-10 \\
M22 & Manager, Female & 33 & 5--10 years  & UX/UI Design Team (Technology)              & 10-15 \\
W23 & Worker, Female  & 27 & 3--5 years   & UX/UI Design Team (Design)                  & 5-10 \\
\bottomrule
\end{tabular}
\end{table*}

Our sample included 13 managers (M2, M4, M5–6, M9–10, M13, M16–19, M21–22) and 10 workers (W1, W3, W7–8, W11–12, W14–15, W20, W23). Participants ranged in age from 24 to 45 years across sectors. Notably, some workers in our sample had relatively long professional experience, while some managers were comparatively early in their careers. This reflected the fact that managerial status was not strictly tied to years of employment but to the nature of participants’ current responsibilities. In several fast-paced industries represented in our sample, such as internet  and media, employees may become team leads or middle managers within one to two years, especially in organizations where project-based coordination or client-facing decision-making is delegated early. By contrast, we also included managers with extensive work experience (e.g., M18, M19, M21, M22), whose teams operated within larger or more hierarchical structures; in these settings, managerial roles emerge through tenure, accumulated expertise, and formal promotion systems. Our dataset therefore reflects a wide spectrum of managerial trajectories, from early-career managers to highly seasoned leaders, as well as workers whose long-term expertise does not necessarily translate into supervisory roles.

In this context, “work experience” refers to total years in the workforce, not the number of years spent in managerial positions. This distinction helps explain why some participants with relatively short work histories were nonetheless categorized as managers, they held formal authority in their current roles, while some participants with longer histories remained individual contributors because their duties did not involve supervising others. For this reason, role categorization in our study followed participants’ self-identification based on formal position and everyday responsibilities rather than tenure alone.

To further strengthen diversity, we also recruited participants from teams of varying sizes, ranging from small groups of fewer than five members to large teams of more than fifty. Including participants from this range allowed us to capture how the imagined role of Manager Clone Agents might differ across organizational structures and accountability demands.

During the workshops, participants were asked to fully inhabit the managerial or worker roles they currently held within their organizations, drawing on their lived experiences to co-create and reflect on speculative futures involving Manager Clone Agents. This emphasis on role-grounded participation ensured that our discussions and scenarios remained contextually rich and sensitive to the diverse ways managerial authority and worker agency are enacted across contemporary workplaces.

\subsection{Design Fiction Workshops}  

Each design fiction workshop lasted approximately 120 minutes, following a structured sequence of activities designed to elicit both imaginative and critical reflections on Manager Clone Agents in future teams and organizations. Aligned with design fiction and speculative design traditions in HCI~\cite{dunne2024speculative, wong2018speculative}, we convened managers and workers in a single workshop and created room for debate about socio-technical values and power dynamics. To offset the power asymmetries between managers and workers, we ensured that each participant had dedicated turns to speak during group activities, and rotated the speaking order of managers and workers in each speaking round. In addition, to reducing the likelihood that managerial voices would dominate, we explicitly encouraged workers in weaker positions of power to comment on or challenge the managers’ concepting. IRB approval was obtained prior to the workshops; all participants provided informed consent, and each received \$30 gift card after participation.

\subsubsection{Introductions and role grounding.} At the start of each session, participants introduced themselves by describing their primary responsibilities within their current organization, their experiences of teamwork, and most importantly, their experiences of managing others or being managed. To anchor discussion in lived realities, each participant was also asked to share three to five concrete examples from their own work where managerial presence or absence significantly shaped outcomes.  

\subsubsection{Scenario brainstorming.} Building on these examples, participants individually brainstormed how their cases might unfold differently if a manager had been replaced by a Manager Clone Agent. Managers were asked to adopt the perspective of a leader delegating authority to a Manager Clone Agent, while workers were asked to imagine themselves being directed by such an agent. This activity served as a recall scaffold that helped participants surface tacit power dynamics and conflicts to later reconfigure in fiction.  

\subsubsection{Design fiction writing.} Each participant then selected one scenario to develop in greater depth, but was also free to expand creatively and develop entirely new directions. They spent 15 minutes writing a short design fiction narrative describing how future workplace practices and relationships would unfold when Manager Clone Agents were involved. We prompted them to stretch or exceed present-day technological assumptions whenever their stories required it. Participants were encouraged to build directly on the examples they had shared earlier, but were also free to expand creatively and develop entirely new directions, including resistant or structurally disruptive possibilities. This balance ensured that narratives remained grounded in lived experience, while allowing space for imaginative exploration of alternative future possibilities. It is noted that these design fictions are based on the imaginaries of our participants, instead of existing Manager Clone Agents deployments.

\subsubsection{Story sharing and role-play discussion.} After writing, the participants took turns presenting their stories to the group. Following each presentation, other participants engaged in a role-play discussion by responding as if they were members of that scenario, either managers or workers based on their own roles, reflecting on how they would experience and react to the presence of a Manager Clone Agent. This process encouraged participants to inhabit perspectives beyond their own and to surface tensions between managerial and worker vantage points.  

\subsubsection{Final round-table discussion.} Once all individual stories and discussions were completed, the group reconvened for a round-table discussion. Building on the scenarios they had created, participants reflected on several guiding questions. First, they considered what kinds of work or duties Manager Clone Agents could reasonably take over on behalf of managers, and which responsibilities should remain human-led. Second, they debated the benefits and concerns of using clone agents for managers, and what implications these might have at the individual (for both workers and managers), team, and organizational levels. Third, participants discussed how Manager Clone Agents should be designed in the future, including what forms of support would be most useful and what boundaries should be preserved. Finally, they compared how managers and workers viewed these issues differently, highlighting the distinct interests and considerations that shaped each group’s stance. These round-table conversations allowed participants to move from individual stories to broader reflections, surfacing shared opportunities, points of contention, and open questions about the integration of Manager Clone Agents in future workplace.  

\subsection{Data Analysis}

We analyze the workshop materials using reflexive thematic analysis \cite{braun2021one, braun2019reflecting}. The dataset included participants’ written design fiction stories, notes from scenario brainstorming, and transcripts from group discussions and round-table synthesis. All workshops were audio-recorded and transcribed verbatim. To preserve the richness of participants’ accounts, we treated fictional narratives and reflective commentary as complementary forms of data: stories articulated imaginative possibilities, while discussions revealed participants’ interpretations, concerns, and reasoning about those imagined futures.

Our analytical approach was primarily inductive. Initial codes were generated bottom-up from participants’ narratives and discussions rather than derived from predefined theoretical categories, while still being informed by sensitizing concepts from CSCW/HCI research. This orientation allowed us to remain open to unexpected insights while ensuring that interpretations were grounded in participants’ meaning-making.

The analysis proceeded iteratively. First, two researchers independently read through all transcripts and narratives to gain familiarity with the data and note initial impressions. Second, we conducted open coding at the sentence and paragraph level, tagging both descriptive content (e.g., tasks automated, benefits cited, risks identified) and relational dynamics (e.g., authority, trust, recognition). Third, we collaboratively grouped codes into higher-level categories that captured recurring patterns across workshops. Throughout this process, we repeatedly returned to the raw data to ensure that emerging categories remained grounded in participants’ accounts.

\begin{figure*}
    \centering
    \includegraphics[width=1\linewidth]{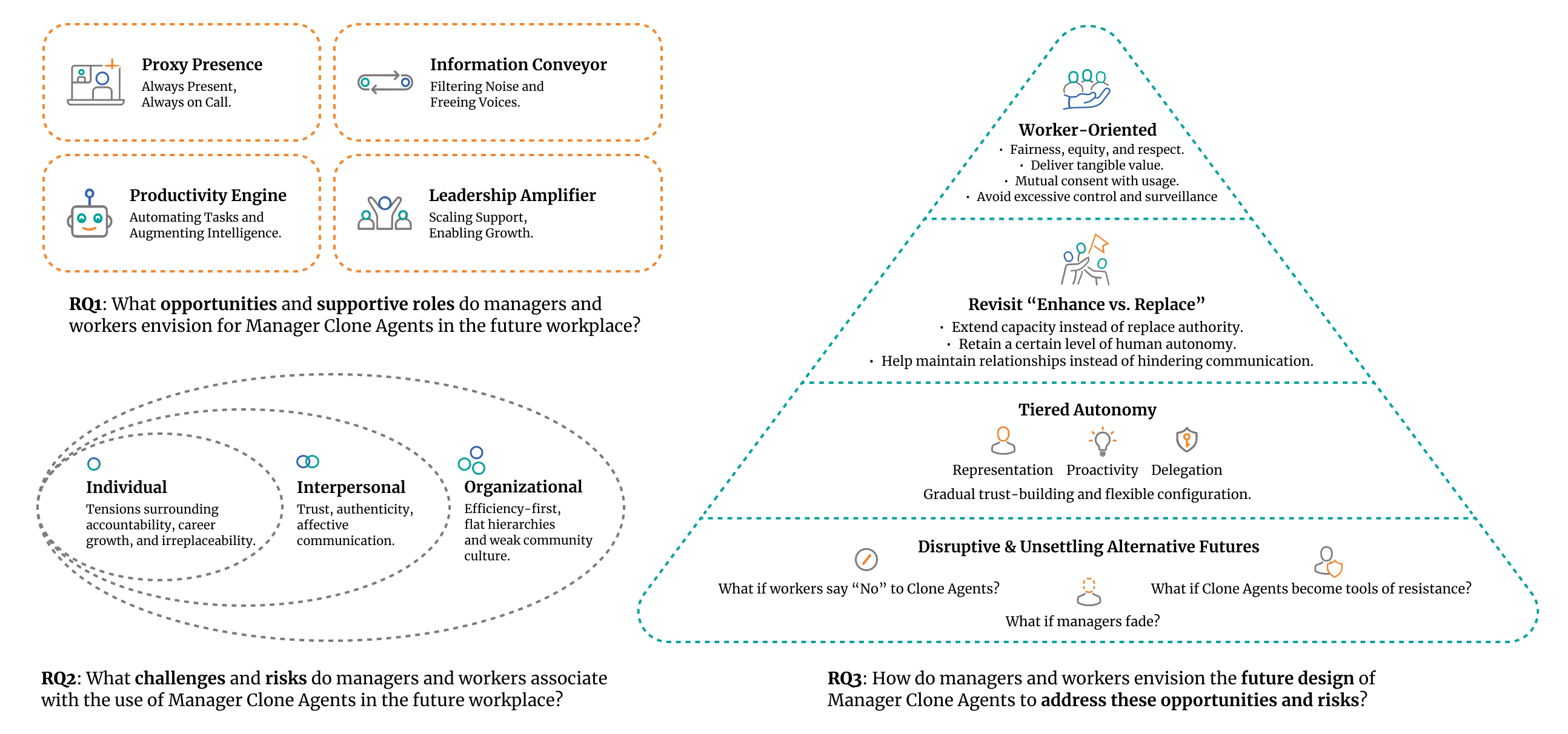}
    \caption{Summary of Findings.}
    \Description{Conceptual framework of findings organized around three research questions. For RQ1 (opportunities), managers and workers envision four supportive roles for Manager Clone Agents: proxy presence (always present/on call), information conveyor (filtering noise and freeing voices), productivity engine (automating tasks and augmenting intelligence), and leadership amplifier (scaling support and enabling growth). For RQ2 (challenges/risks), anticipated tensions cluster at individual (accountability, career growth, irreplaceability), interpersonal (trust, authenticity, affective communication), and organizational levels (efficiency-first logics, flat hierarchies, weak community culture), shown as overlapping layers. For RQ3 (design considerations), a tiered pyramid highlights worker-oriented principles (fairness/equity/respect, tangible value, mutual consent, avoiding excessive control/surveillance), a call to revisit “enhance vs. replace” (extend capacity without replacing authority, retain human autonomy, support relationships), and tiered autonomy (representation → proactivity → delegation, with gradual trust-building and flexible configuration). The base emphasizes disruptive and unsettling alternative futures, posing critical “what if” scenarios (workers refusing clones, managers fading, and clones becoming tools of resistance) that shape future workplace visions.}
    \label{fig:findings}
\end{figure*}

The final stage involved synthesizing categories into three overarching themes that structured our findings (see~\autoref{fig:findings}) : (1) the supportive roles participants envisioned for Manager Clone Agents (see \autoref{sec:RQ1}), (2) the risks and tensions anticipated at individual, interpersonal, and organizational levels (see \autoref{sec:RQ2}), and (3) the future design directions participants proposed (see \autoref{sec:RQ3}). Although the themes align with the study’s research questions, this alignment reflects the shared grounding of both the questions and the themes in participants’ accounts. The research questions framed the scope of inquiry, but the themes themselves emerged inductively through the bottom-up clustering of categories rather than from a predefined coding frame.

Design fiction played a central role in shaping our analytic process. Because participants’ narratives intentionally exaggerated, extended, or reconfigured familiar workplace dynamics, the fictional mode acted as a probe that surfaced latent assumptions about authority, collaboration, delegation, and managerial presence. During coding, we treated fictional exaggerations not as predictions but as windows into participants’ organizational imaginaries, enabling us to identify concerns and relational patterns that may remain unarticulated in conventional interview settings.

To enhance trustworthiness, the research team held regular debriefing sessions to challenge assumptions and refine interpretations. Disagreements were resolved through discussion and multiple iterations of coding and theme development were used to balance consistency with openness to new insights. Consistent with qualitative HCI practices (e.g., \cite{xiao2025might,fan2025user}), we did not assess statistical reliability; instead, we emphasized interpretive depth, contextual grounding, and transparency in the analytic process.

\section{Findings (RQ1): Empowering Teamwork: Four Potential Supportive Roles of the Manager Clone Agents} \label{sec:RQ1}

Our findings reveal that participants envisioned Manager Clone Agents to empower future teamwork across four distinct supportive roles. As \textit{a Proxy Presence}, they were expected to mitigate the costs of managerial absence by sustaining responsiveness and symbolic availability when managers were busy and not available (see~\autoref{sec:proxy}). Acting as \textit{an Information Conveyor}, they were imagined to streamline communication flows, reduce distortion, and surface grassroots concerns that might otherwise be lost in hierarchies (see~\autoref{sec:belt}). As \textit{a Productivity Engine}, they were expected to expand managers’ capacity by automating routine tasks while also freeing cognitive bandwidth for more strategic work (see~\autoref{sec:engine}). Finally, as \textit{an Amplifier of Leadership}, they were envisioned to extend the manager’s coaching, feedback, and mentoring to a broader set of workers, compensating for the limits of managerial time and attention and magnifying guidance that would otherwise remain scarce (see~\autoref{sec:amplifier}).

\subsection{A Proxy Presence that Preserves Managers' Continuity and Symbolic Authority}\label{sec:proxy}

Participants consistently described Manager Clone Agents as forms of proxy presence that could stand in for leaders when physical attendance was difficult or impossible. They envisioned these agents not as full replacements, but as a way to project presence, maintain communication, and preserve authority across different situations. This proxy presence was articulated in three ways: first, as a bridge during crises when managers could not be present; second, as a tool for more flexible and asynchronous participation in everyday work; and third, as a symbolic extension of managerial authority that differed in meaning from delegating to human assistants.

Many managers and worker participants in our study emphasized that the burdensome nature of leadership responsibilities often makes physical presence difficult. M10 described how managers like him are constantly on the move: \textit{“meeting clients, traveling across regions, and balancing competing responsibilities.”} He added that this constant mobility left him mentally and physically exhausted, as he was expected to maintain visibility and responsiveness while being pulled across multiple sites of work. Other managers echoed this sentiment, noting that such demands delayed responses and reduced availability. Several workers, including W8, W12, and W15, also described feeling hesitant to approach managers who were perpetually “\textit{elsewhere},” worrying that asking for help would add to their burden.

In this context, participants saw Manager Clone Agents as a way to reduce the social costs of absence by providing a form of presence when managers could not attend in person. For example, many worker participants think, rather than leaving a meeting empty or canceled, the agent could appear with a simulated voice and avatar, signaling the manager’s involvement in a more respectful way than a brief apology about the disappearance.

M17, a marketing manager with 5–10 years of experience, illustrated this possibility in the speculative design story she wrote. In her scenario, a family emergency coincided with an important cross-timezone stakeholder meeting. Rather than canceling, she imagined sending her avatar, having pre-uploaded key talking points so that “\textit{the discussion could still move forward even if I wasn’t there in person}.” 

Beyond crisis situations, participants also saw Manager Clone Agents as tools for more flexible participation. By asynchronously mediating communication, they allowed both managers and workers to engage at their own pace rather than in real time. W15 explained: \textit{“With the Manager Clone Agent standing in, it’s the same as leaving a voice message. I can just listen whenever I have time, until it’s the best time for me.”} 

Additionally, compared with interacting directly with the real manager, some worker participants described feeling less pressure to give undivided attention when meeting with clone agents. As W15 put it, \textit{“If they’re talking about results, I’ll just leave it on in the background, like a Zoom meeting with camera and mic off.”} This comment highlighted how the mediated presence of a Manager Clone Agent shifted expectations: workers could still access managerial input without the same demand for immediate, face-to-face attentiveness. 

Participants also noted that the symbolic presence of a Manager Clone Agent carries different weight than sending an assistant or proxy human. As M21 put it, \textit{“Even though it’s a clone, it’s your leader’s clone, not another person who might distort information.”} Here, the symbolic value lies in the continuity of identity: the clone embodies the manager’s voice and authority, even if mediated, whereas an assistant is seen as a separate actor. M18 believed that, for broadcast-style meetings or official statements, this symbolic link made the clone agent feel more authoritative and representative than delegating to other staff. At the same time, participants emphasized that symbolic presence has limits. As M6 noted, the manager’s personal appearance still matters in high-stakes contexts where trust cannot be fully replaced by a clone.

\subsection{An Information Conveyor that Clarifies Directives and Surfaces Grassroots Feedback} \label{sec:belt}

Participants repeatedly highlighted the communicative role of managers as conduits of information—\textit{“passing directives downward and carrying feedback upward, particularly for mid-level managers,”} as M10 described. They envisioned Manager Clone Agents as impartial conveyors that could transmit policies and feedback more clearly, reduce cognitive misalignment, and create safer channels for workers’ voices to be heard. Beyond individual exchanges, participants also imagined these agents improving large-scale information gathering, amplifying grassroots concerns that might otherwise be lost or filtered.  

In discussing impartial transmission, M18 noted that messages relayed through human intermediaries were \textit{“less likely to be impartial”}, often shaped by bias, incomplete understanding, or social discomfort. Building on this, workers like W7 and W8, along with managers such as M21, imagined Manager Clone Agents as more neutral carriers. In their stories, agents could smooth over gaps, keep meaning intact, and reduce the confusion that often arises when policies and feedback travel through layers of hierarchy.  

Participants also reflected on how Manager Clone Agents might reduce friction caused by cognitive misalignment between managers and workers. W8 recounted the difficulty of explaining her visa situation to a manager unfamiliar with international work rules: \textit{“Earlier I discussed with my manager how I could extend my OPT. Since he is not an international worker, I needed to provide him with some context and continue to tell him what my timeline looks like, and he has to continue to report to his manager and human resources.”} In her view, the problem was not unwillingness but the sheer effort required to keep restating context across different levels of hierarchy. A Manager Clone Agent, participants suggested, could retain and transmit such contextual knowledge on behalf of the worker, ensuring that details are not lost each time the message is passed upward, while also sparing workers from the exhausting cycle of repeated clarification.

Another recurring theme was the potential to free voices that might otherwise be silenced by hierarchy. M18 observed that \textit{“in the workplace, workers may have some requests and not dare to say them directly.”} W20 in the same workshop echoed this, imagining Clone Agents as mediators that could create safer channels for raising concerns. In one speculative story, M4 described recruitment scenarios where applicants could ask sensitive questions more comfortably to an agent than to a manager, since the agent could respond with patience and without judgment, and applicants would not feel embarrassed or disadvantaged for raising such issues. 

Finally, participants described how Manager Clone Agents might improve large-scale information collection. M21 imagined an agent preparing for a general assembly by first meeting with smaller groups to surface pressing issues: \textit{“To prepare for a general assembly, an Agent could meet with smaller groups (departments, floors, etc.) to figure out which pressing issues should be addressed.”} W23, in the same workshop, agreed that this would make workers feel heard, contrasting it with surveys: \textit{“If you have the Manager Clone Agent, you will feel like someone is listening to you and they will do something about it.”}

\subsection{A Productivity Engine that Prevents Fatigue and Bias, Freeing Leaders for Strategic Work} \label{sec:engine}  

Participants expected Manager Clone Agents to improve productivity in two directions. On the one hand, agents can take over repetitive, rule-based work and prevent mistakes that come from human limits such as personal preference, fatigue, or bias. On the other hand, agents can strengthen managers’ decision quality by remembering more context and giving more consistent recommendations. This combination would shift managers’ effort away from low-level tasks toward strategic and relational responsibilities.

Many participants described how Manager Clone Agents could absorb what M13 called \textit{“high-frequency, low-risk, and rule-based”} processes that otherwise drained managerial time and attention. Routine approvals, such as \textit{“approving invoices”} (M6), \textit{“verifying expense claims”} (M13), \textit{“processing leave”} (W1, M10), or \textit{“signing off on weekly timesheets”} (W12), were repeatedly mentioned as tasks that could be offloaded.

Agents were also expected to prevent performance drop-offs caused by human bias. M21 noted that \textit{“people may judge, mislead or distort information, whereas a system is imagined as less prone to such flaws”}. M5 described that \textit{“leaders interviewing many candidates can lose focus and forget earlier details; an agent’s steadier attention would catch inconsistencies that humans miss when attention is not concentrated”}.

Beyond avoiding failure, agents were expected to push performance beyond human capacity. Participants emphasized machine strengths like \textit{“deeper contextual memory”} (M19) and instant access to \textit{“domain or internet knowledge”} (W11). M19 suggested that \textit{“a CEO’s repeated all-hands or cultural onboarding could be delivered by an agent who might speak even better than the CEO, freeing leaders from repeating the same material”} (M19). W11 similarly imagined agents searching for online information rapidly to improve proposal revisions when leaders are busy or outside their expertise (W11).

Finally, by compressing details into quick takeaways, agents could let leaders \textit{“review in five minutes and spend time on higher-level judgment and abstract decisions”} (M9). But participants also warned that misalignment between agent actions and leader intent can create coordination costs or reversal risks. M2 feared that \textit{“even small deviations could lead to ‘180-degree’ execution swings and wasted meetings if the leader overturns the agent later”} (M2). Others noted that \textit{“agents still miss human grayspace and tacit intentions in scheduling or tradeoffs”} (W8).

\subsection{An Leadership Amplifier that Scales Mentorship and Allows for Autonomy and Growth} \label{sec:amplifier}  

In our workshops, participants envisioned how Manager Clone Agents might amplify a manager’s impact by scaling day-to-day guidance to multiple workers at once, widening access to the manager’s attention, and creating space for workers to develop independent judgment.

Participants described Manager Clone Agents as useful for coordinating work and giving timely guidance so that tasks do not stall while waiting for a meeting. In W15’s story, the agent greeted him on Monday with a ranked task list, answered common questions, and, when reaching its limits, pinged the manager while suggesting productive interim tasks. M9 similarly imagined an agent from his supervisor accelerating turnaround times, while M2 described lightweight nudges such as a Slack bot that periodically reminds workers of handoffs or check-ins.  

Amplification also meant making managers more accessible to workers. W7 explained that a Manager Clone Agent could capture feedback right after a meeting that might otherwise be lost if the real manager was traveling or tied up in back-to-back calls. W8 added that reaching an agent felt more approachable and responsive than sending what she called a “\textit{cold email}.”  

Managers stressed that this kind of amplification works only when agents enforce role boundaries and escalate issues at the right thresholds. M16 envisioned setting clear pre-approved decision bands in her story, noting that \textit{“My team can ask my agent if I am okay with certain decisions or then alert me if my authority or insight is needed for a potential project decision. My team would view this as potentially having greater access to me and the resources to get the job done. They would be able to feel more empowered to make decisions and could run thoughts by me more quickly.”} At the same time, W23 cautioned that \textit{“There’s a difference between giving someone feedback and spoon-feeding them, the latter would create a reliance on constantly having someone to bounce ideas off 24-7.”} If the agent is always available, juniors may defer decisions they should learn to own.  

Finally, participants described softer forms of amplification, such as carrying culture and care. M9 imagined a Manager Clone Agent that “\textit{facilitates focus sessions and mirrors a mentor’s way of thinking on a mentee’s slides or writing}”, thereby “\textit{passing on both ideas and spirit}”. M18 shared how an automated birthday note from a system unexpectedly rekindled a relationship with a former teammate; others saw similar low-stakes, high-touch gestures as well-suited for Manager Clone Agents to maintain warmth and connection across large organizations.

\section{Findings (RQ2): Navigating Risks: Individual, Interpersonal, and Organizational Impacts}
\label{sec:RQ2}

While the previous section highlighted how participants envisioned opportunities for extending managerial reach, here we turn to the risks that, in our participants’ imaginaries, could emerge if Manager Clone Agents were poorly integrated into everyday practices. These risks spanned individual, interpersonal, and organizational levels. At the individual level (see~\autoref{sec:individual}), both managers and workers grappled with shared vulnerabilities around accountability, skill erosion, and replaceability, though their stakes diverged: managers feared threats to authority and identity, while workers worried about blocked career pathways and recognition. At the interpersonal level (see~\autoref{sec:interpersonal}), trust initially transferred from human relationships to Manager Clone Agent-mediated exchanges, but fragile authenticity, weakened affective cues, and reduced informal contact strained workplace bonds. At the organizational level (see~\autoref{sec:organizational}), efficiency gains were anticipated to flatten hierarchies and accelerate workflows; yet, participants feared this would hollow out belonging and weaken culture.

\subsection{Individual Risks: Divided Titles, Shared Vulnerability} \label{sec:individual}  

Across discussions, managers and workers found themselves bound by similar anxieties yet separated by divergent stakes. Both recognized that Manager Clone Agents unsettle the rules of responsibility, reshape opportunities for growth, and test what it means to remain irreplaceable for human employees. But while managers viewed agents as a challenge to their authority and professional identity, workers saw them as blocking pathways to recognition and advancement. These overlapping perspectives reveal how Manager Clone Agents use exposes both shared vulnerabilities and role-specific divides. 

Participants often returned to the question of accountability, disagreeing on who carries the blame when a Manager Clone Agent speaks or acts. Concerns centered on errors: agents misinterpreting intent, causing harmful outcomes, or introducing communication mistakes. Managers worried that outputs might drift from their “real” intent because tacit context cannot be cloned, with M18 explaining: \textit{“You can’t collect a lifetime of experience into a knowledge base, so some things the agent says won’t be what the leader truly wants”}. Others recalled being held responsible for actions they had not authorized in their design fictions, as M10 noted: \textit{“You come back and find you’ve carried a lot of blame for things your agent said.”} Workers, however, resisted being scapegoated. W3 asked, \textit{“If my AI made a decision that later caused problems, is the blame on me, on the AI, or on whom?”} and W12 added, \textit{“If I revise work per the agent’s advice and it backfires, I’m probably the one who takes the fall.”} Participants also anticipated \textit{“misunderstandings from transmission errors”} (W1) and worried that agents might become a convenient excuse for poor human stewardship: \textit{“It gives people something to blame”} (W7). As M16 summarized, \textit{“If they lean on my avatar and the output becomes distorted, and no one owns it, accountability becomes the issue.”}  

Participants also framed Manager Clone Agent use in terms of career opportunities, but from opposite vantage points. Some managers such as M2 and M10 imagined reinvesting time saved into managing up and outward: cultivating senior relationships, doing market visits, or expanding scope. Workers in the same workshops described the flip side: sparser access to managers, feeling devalued, and losing motivation. As W3 put it, \textit{“I kind of just don’t really feel like working anymore.”} For workers, even routine updates or casual chats with supervisors were seen as crucial opportunities to demonstrate impact. W12 explained: \textit{“Weekly updates to the boss and some casual chat are themselves an opportunity to build rapport and leave a positive impression. You need to show your manager what impact your work has had, if you want a promotion.”} Managers, meanwhile, feared that over-delegation to clone agents would cause skills to atrophy. As M4 warned, \textit{“People are greedy, the more they use it, the lazier they become.”} To hedge this risk, others stressed the need to preserve core judgment for managers, as M16 insisted: \textit{“I want to keep control over my analysis of the metrics. AI can suggest targets, but I still set them.”}  

Participants further debated the extent to which managerial roles themselves might be replaced. Some workers argued that project managers with limited decision-making authority were highly vulnerable. W12 estimated \textit{“I think 80\% of his work could really be replaced by an agent.”} W3 similarly questioned the uniqueness of managerial labor, remarking: \textit{“I don’t think the division of labor made by AI or real managers will be any different, because I feel that neither is perfectly fair.”} 

Many managers such as M2, M5, M6, M10, M13, M18 and M19 defended higher-order decision-making, creative judgment, and relational work as difficult to automate. Yet some still admitted that agents might still perform adequately to some extent. As M17 reflected, \textit{“I would just have to resign to like it just wouldn’t be done maybe the way that I would do it. And that doesn’t mean it’s wrong. It can still do good enough.”} Such reflections pushed managers to confront what M17 called \textit{“existential and ego connected identity questions,”} echoed by M9: \textit{“I probably don’t want to authorize him because if my own work could be done by my agent, then what is this job position for me to do?”}

\subsection{Interpersonal Risks: Trust Cloned, But Wears with Distance}
\label{sec:interpersonal}
Trust in Manager Clone Agents was initially built on trust in the human managers they represented. Participants explained that because the clone was seen as an extension of a known person rather than an anonymous system, the trust cultivated through prior human relationships could be partially transferred to the agent. In this sense, the “cloned” trust was borrowed credibility: workers felt more comfortable engaging with a Manager Clone Agent when they already trusted the manager behind it. Yet participants also emphasized that this form of trust was fragile. Unlike trust in a human being, which is reinforced by lived interactions, judgment, and emotional reciprocity, cloned trust rested on the assumption that the agent would faithfully mirror the manager’s intent and style. Once communication became mediated by agents, this assumption was tested and often strained by misinterpretations of value, uncertainty about authenticity, and the loss of affective and informal cues. 

Participants first highlighted that the credibility of Manager Clone Agents depended on existing human relationships. M21 explained that \textit{“people will be more open with Manager Clone Agents than generic agents as it’s talking to someone you know vs. someone you don’t know.”} Without this \textit{“underlying relationship,”} M16 argued, the system would not constitute a \textit{“good functioning system.”} 

Participants then noted that trust weakened when managers relied too heavily on agents for direct interactions. They emphasized that agents could be useful to enhance communication, such as speeding up responses, filling gaps when managers were unavailable, or relaying straightforward information. But when small-group or one-on-one conversations were fully delegated to a clone, the meaning changed. Workers felt devalued, interpreting it as a signal that their work was \textit{“unimportant”} (W3, W15) or that they were \textit{“unworthy of their manager’s time”} (W14). This perception, M9 warned, could \textit{“diminish motivation.”} W14 expressed the frustration more bluntly: \textit{“If you can’t be bothered to work with me at all, then why would I respect you?”} Managers also worried about this dynamic in reverse, interpreting workers’ reliance on agents as avoidance or disengagement. As M6 explained in another perspective, \textit{“Managers wouldn’t like it if workers use a partition layer to block the imagined harm the boss might inflict on them.”} In short, participants saw Manager Clone Agents as acceptable supplements to human communication, but resisted their use as substitutes for genuine interaction.

\begin{figure*}
\centering
\includegraphics[width=0.75\linewidth]{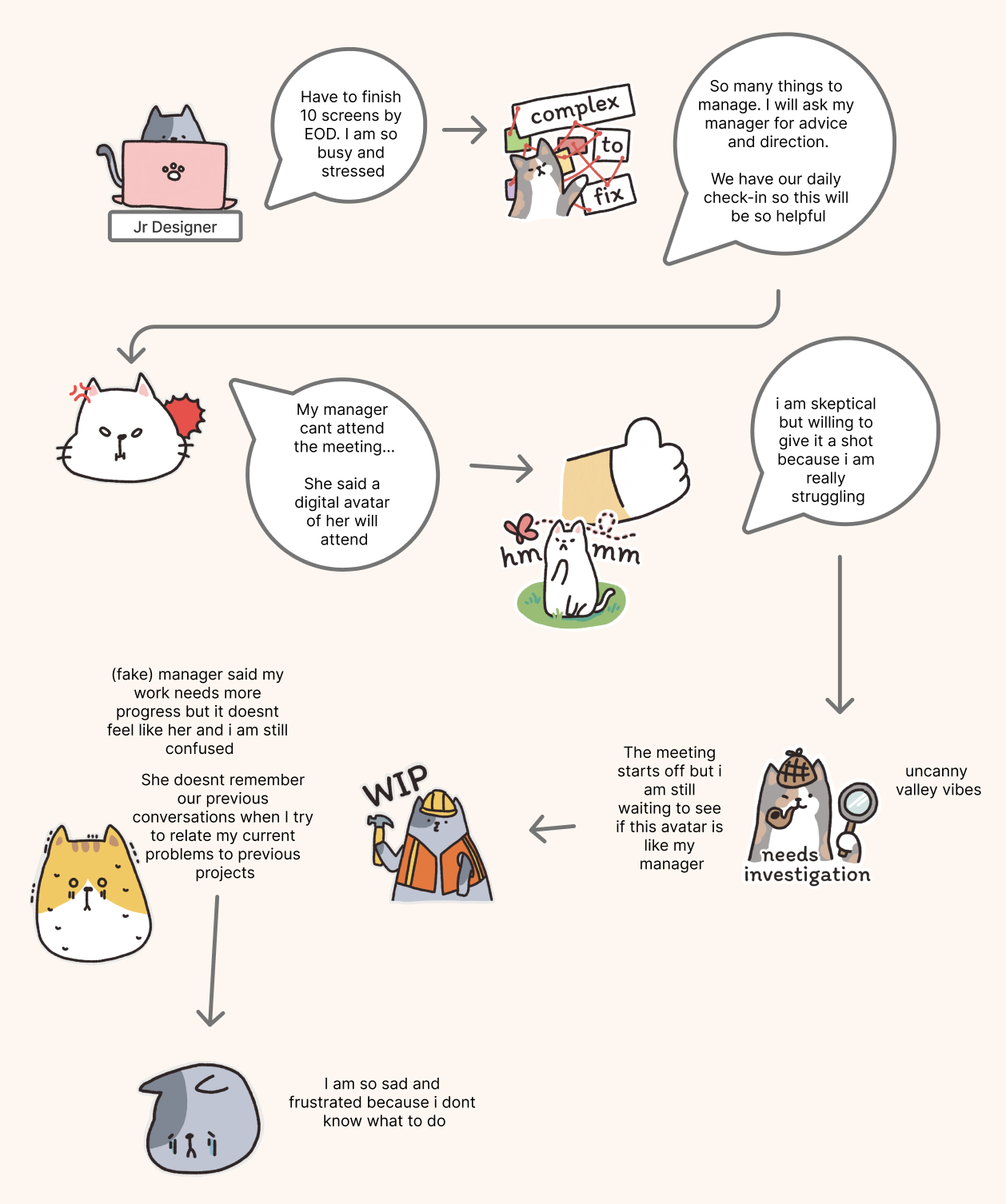}
\caption{Story From W23 Showing Frustration When the Manager Clone Agent Fails to Recall Past Conversations.}
\Description{A visual related to W23’s Story on Decision-making and administration, highlighting the confusion caused by overlapping identities and the frustration that arises when the agent fails to remember past conversations between real individuals.}
\label{fig:placeholder}
\end{figure*}

Trust was further strained by uncertainty over authenticity. Workers such as W3 and W14 said they often could not tell whether a message came from their manager, from the agent itself, or whether their own contributions were being faithfully conveyed. Because Manager Clone Agents were designed to stand in for real people, this ambiguity directly undermined the very premise of the clone: workers needed to believe it faithfully mirrored their manager, yet could not be sure where the manager ended and the agent began. W23 illustrated this confusion on a storyboard (see \autoref{fig:placeholder}), showing the frustration that arises when overlapping identities blur accountability and when agents fail to recall past conversations between real individuals.

Participants also emphasized the loss of emotional nuance and informal connection. W3 acknowledged that one benefit was the absence of negative outbursts, since an agent would not lose its temper. Yet even when agents simulated appropriate affect, workers often found it \textit{“unconvincing or even demotivating”} (W1). The absence of genuine cues reduced moments of casual bonding. As W3 described, \textit{“At the beginning of meetings, there is usually a warm-up stage where we chat casually. I think this is when connections between people are built, and that gets lost interacting with agents.”} This point sparked debate: M4 felt that authentic bonds were created outside formal settings such as meals or after-hours chats, while W3 considered meeting warm-ups essential. Managers also imagined longer-term erosion of interpersonal ties. M2 wrote in his design story a horrible relationship development that \textit{“The number of times I communicate with my staff has decreased, and to me their work is just a schedule sheet on a screen. I can’t even recall the names and appearances of newly hired employees anymore. The anonymous subordinate satisfaction feedback from HR is getting lower and lower.”}  

Participants stressed the need to balance efficiency with the preservation of interpersonal connection. As M16 remarked, \textit{“If I was leading a team, I’d want to make sure that personal connection was still there and that they don’t feel as if they’re just being offloaded.”} In their view, trust that is cloned through Manager Clone Agents risks wearing away with distance, gradually reshaping the texture of workplace relationships and, eventually, organizational culture (see~\autoref{sec:organizational}).

\subsection{Organizational Risks: Performance Up, Belonging Down}
\label{sec:organizational}

At the organizational level, participants described a paradox. They anticipated that Manager Clone Agents could improve systemic efficiency: reducing layers of coordination, redistributing managerial responsibilities, and broadening access to leadership. At the same time, they worried that these very efficiencies might erode the social fabric of organizations, harming inclusiveness, and reshaping collective identity. In other words, while Manager Clone Agents promised to streamline structures, participants feared that they might hollow out the cultural role that workplaces have long played.  

The participants highlighted efficiency as a powerful driver of structural transformation. M18 described how Manager Clone Agents could take over repetitive approvals and routine tasks, relieving managers of \textit{“annoyances”} and freeing up \textit{“an extra hour or two every day.”} W12 imagined that if such delegation became common, \textit{“maybe there just aren’t that many managers anymore, the organization is becoming increasingly flat.”} Echoing this, M18 noted that many intermediary roles such as HR staff or project managers exist mainly \textit{“to facilitate communication between managers and front line workers,”} positions that agents could absorb. While some participants welcomed this flattening as a path to accessibility and speed, others feared: “\textit{This will cause the loss of competition that historically drove innovation”} (M10). 

Apart from the power structure, participants warned about the risks of a fractured culture within the organization community. W1, pointed to inequities that might emerge: “\textit{Workers willing to rely heavily on agents could offload routine labor and free up time for other pursuits, while those unwilling or unable to adopt the technology would be forced to carry out more tasks themselves}.” She envisioned: “\textit{Over time, such asymmetries could create divisions between workers’ groups and amplify perceptions of unfairness}.” M2 dramatized this erosion in a story where “\textit{dissatisfied employees were gradually replaced by those who no longer cared}”. W3 pushed the scenario further, imagining a future where \textit{“The employees of the entire company are all working for AI,”} leaving communal identity hollow. M9 satirized the endpoint: “\textit{An office filled entirely with mobile AI agents while humans stayed home: some enjoying more family time, others left isolated}.” 

\section{Findings (RQ3): Designing for Manager Clone Agents in Future Workplace}
\label{sec:RQ3}

Having surfaced both hopes and anxieties around Manager Clone Agents, participants turned toward imagining how such systems should be designed and deployed. Three interconnected design orientations emerged: keeping workers’ needs at the center (see~\autoref{sec:worker}), positioning Manager Clone Agents as supportive extensions rather than replacements for managers (see~\autoref{sec:enhance}), and structuring agents along graduated levels of autonomy to allow context-sensitive use and gradual trust-building (see~\autoref{sec:autonomy}). Alongside these orientations, participants also ventured into more disruptive speculative imaginations that unsettled the assumed permanence of managerial roles, pointing toward alternative futures in which workers resist, managers fade, or clone agents are repurposed in unexpected ways (see~\autoref{sec:alternative_future}). 

\subsection{Worker-Oriented: Supporting Workers Rather Than Surveilling Them}
\label{sec:worker}

While managers may see such systems as convenient extensions of their authority, workers are the ones who interact with them and who directly bear the consequences of managerial clone agents usage. The participants, therefore, pointed out that it is highly important to prioritize the pains and needs of the workers and support their work when designing Manager Clone Agents.

Concerns about excessive control and surveillance were repeatedly emphasized. M19 observed that workers can feel as if they are reporting to “\textit{two leaders},” one being an extremely rational and emotionally detached version of their manager who “\textit{keeps asking endless questions}.” M17, reflecting on her own workplace, stressed that she did not want colleagues to feel as if they were “\textit{always on the spot}” or “\textit{watched the way that some companies just have your camera on all day}.” For her, technology that fosters this feeling amounts to micromanagement. 

Both managers and workers emphasized that their top concern is whether these systems provide concrete benefits to workers; otherwise, they will face pushback. W7, for example, proposed the idea that “\textit{An agent could improve and polish documents before sending them to the manager}”. More broadly, as mentioned in \autoref{sec:RQ1} workers expressed interest in Manager Clone Agents streamlining administrative tasks in team collaboration. In these cases, clone agents were seen as a supportive tool that reduced invisible labor. As M18 put it bluntly, most workers approach their jobs “\textit{just to earn a wage},” and if a digital agent suddenly appeared to “\textit{constantly command}” them, making work “\textit{increasingly difficult}” or adding “\textit{many annoying tasks},” they would “\textit{hate it deeply}.” As M21 summarized: “\textit{workplace dynamics often feel like us versus them, whether the Manager Clone Agents shows that it cares about workers will shape how they respond to it.}”

These tensions further underscore the importance of clarity and consent. Workers pointed out that the legitimacy of Manager Clone Agents depends on explicit agreements about where they may or may not be used. For example, W12, a junior programmer, argued that a one-on-one meeting is a crucial opportunity to “\textit{leave a good impression}” and demonstrate initiative, and substituting an agent in this setting would feel “\textit{disrespectful}.” W15 echoed this sentiment, insisting that in contexts “\textit{where emotional intelligence is important, I want the real person}.” At the same time, he envisioned negotiating expectations directly with his manager: deciding together which interactions could be delegated to the agent and which must remain human-led. 

Yet not all contexts lend themselves to easy consensus. M19, a company Vice President, described how he considered on-boarding sessions fully replaceable: “\textit{It’s repetitive and a waste of time}.” By contrast, W1, who has only worked for 1-2 years, stressed that orientation was precisely when human presence mattered most: “\textit{If I’m a new employee joining a company, there might be many things that need to be coordinated with me, I would prefer a real person to handle that coordination with me}.” Such disagreements illustrate how consent is not only about formal negotiation, but also about uneven power and priorities. Managers may be motivated to offload repetitive tasks, while workers may see those same moments as critical to building relationships and trust. Because not all workers feel adequately empowered to refuse the presence of a Manager Clone Agent, the participants advocated organizational safeguards, such as granting workers the protected right to request “\textit{human-only}” interactions or limiting the contexts in which Manager Clone Agents may operate.

\subsection{Revisit “Enhance vs. Replace”: Bridging the Relationships Rather Than Breaking Them}\label{sec:enhance}

The HCI community has long advocated for AI that enhances rather than replaces human capabilities~\cite{10.1145/3544548.3580959}. Our findings extend this debate by showing how the line between enhancement and replacement is inseparable from the quality of manager–worker relationships. Participants judged the value of Manager Clone Agents less by efficiency alone than by whether these systems strengthened or weakened the bonds between humans. Importantly, we found that sometimes managers and workers often imagined enhancement in different terms, reflecting their divergent values and stakes in organizational life.  

For managers, enhancement meant extending their capacity without threatening authority or identity. M17 expressed her wish for the agent to help her \textit{“be a better manager,”} for example, by reminding her when she was \textit{“over budget”} or helping her track priorities, rather than taking over leadership itself. Similarly, M19 stressed that the agent should remain in a service role, \textit{“approving travel requests when a boss is unavailable”} but not \textit{“always pointing fingers.”} M16 echoed this distinction, noting that she was comfortable when the system was \textit{“clarifying what’s already there,”} but grew uneasy when it began \textit{“creating new responses,”} since that crossed into authorship and identity.  

For workers, enhancement meant maintaining access to the leadership while buffering the potential conflicts. W7 imagined using a Manager Clone Agent to manage emotions before approaching her boss: when she feared she might \textit{“lose control”} in a heated moment, the agent could help frame her message more constructively. W15 described the agent as \textit{“an extension of the manager”} that could provide information or quick decisions while still leaving the escalation to the human manager. For workers, enhancement thus meant greater availability and emotional scaffolding, rather than authority protection.  

Yet the line between enhancement and replacement remained blurry. W8’s story illustrated this dilemma: \textit{“I asked, ‘Then how should I communicate this?’ The Agent replied, ‘Do you want me to pass it on or do you want to do it yourself?’ I said, ‘What do you think?’ The Agent said, ‘If you do it, it can help you practice your communication skills, but you might not get what you want. But if I do it for you, it might be more effective but you wouldn’t be involved in the process.’”} While workers saw this as a trade-off between empowerment and convenience, managers like M5 worried that such mediation of clone agents could \textit{“build an imaginary wall that ends up hurting both sides,”} deepening misunderstanding rather than reducing it. 

Even as participants offered these visions of enhancement, they consistently stressed the importance of retaining human autonomy. M16 argued that preserving creativity and trust within teams is critical, noting that relationships, respect, and psychological safety “\textit{take time to develop}” and cannot be delegated to machines. She warned that if technology undermines “\textit{any version of I can’t trust it, or I don’t know, or I don’t feel comfortable},” it risks eroding the interpersonal foundations that drive productivity. M17 echoed this sentiment by emphasizing the value of choice. She wanted to “\textit{see it all, but be able to say like, I’m not busy right now, but I trust that bot to answer for me or I’ll get back to you in 5 minutes}.” 

\subsection{Tiered Autonomy: Gradual Trust-Building and Flexible Configuration}\label{sec:autonomy}
As participants imagined how Manager Clone Agents might operate in practice, they often turned to the metaphor of “\textit{levels of autonomy}” familiar from self-driving cars (M4, M18). Rather than treating Manager Clone Agents as a single, fixed entity, they described autonomy as a spectrum, structured along dimensions such as representation, proactivity, and delegation. These dimensions offered a shared language for debating what forms of cloning would feel acceptable, what boundaries should be respected, and how much authority agents ought to exercise in different workplace contexts.

\subsubsection{Representation.}
Participants emphasized the spectrum of representational fidelity, ranging from highly stylized, cartoon-like avatars to hyper-realistic deepfakes indistinguishable from the human manager. For modality, at the lowest level, Manager Clone Agents may exist as text-based entities; at higher levels, they may incorporate voice, video, or even embodied presence in virtual or physical environments. M22 expressed discomfort with Manager Clone Agents that “\textit{perfectly replicated their speech patterns and idiosyncrasies}”, finding the imitation “\textit{uncanny and inauthentic}”; workers similarly noted heightened anxiety when they could not discern whether communication originated from the manager or the agent (W3, W14). M5, W14 and W15 agreed that abstract forms of representation, such as “\textit{cartoon styled} (M5) or “\textit{orb-like}” voice interface (W14, W15), were perceived as less deceptive and more broadly acceptable. W15 further explained that embodiment affects comfort: “\textit{If it’s an avatar, it requires more attention, more effort to be present, like I’m performing extra socially}.” Text-based agents, on the contrary, were viewed as lightweight and flexible, more easily integrated into everyday practices such as Slack channels.

\subsubsection{Proactivity.}
Another dimension concerns the level of initiative that Manager Clone Agents should exercise, which participants saw as varying by task. At the lowest level, agents might simply observe or log interactions without intervening. At intermediate levels, they could relay pre-authorized statements or provide clarifications. The highest levels would involve spontaneous contributions or even steering discussions that many participants found unacceptable for sensitive or high-stakes contexts. As noted in \autoref{sec:enhance}, M16 and M17 worried about being misrepresented by agents making statements they had not endorsed, or by discussions diverted in their absence in ways that later shaped decisions. M18 cautioned that disclosure must be carefully limited, warning, \textit{“The authority to disclose information should actually be tightened as well. For example, if it involves company finances and someone says that cash flow is tight, workers might panic.”} Similarly, M19 argued that strong presence could backfire sometimes, suggesting, \textit{“Don’t make their presence too strong. Even just occasional reminders or emotional support are enough to help everyone gradually accept this new colleague. If their presence is too strong, everyone will feel threatened.”} 

\subsubsection{Delegation.}
Finally, the scope of action delegated to Manager Clone Agents emerged as a central concern. Low-delegation configurations would confine agents to communication support, while high-delegation systems could independently execute managerial tasks, such as reallocating resources or authorizing workflows, without human intervention. M13 stressed limits: “\textit{As a strictly constrained ‘assistant,’ emphasize ‘reminders’ rather than ‘decisions,’ with transparency and explainability prioritized}.” M18 described the risks of overstepping: “\textit{Some overly specific suggestions should be constrained. The remaining decision space should be left to the person doing the work.  Otherwise, it’s like having a CTO constantly coaching right next to the coder, which makes workers uncomfortable}.” M9 proposed explicit labeling: “\textit{Label the level of an authorized agent, for example, before getting a Gold level no one pays attention, but after getting Gold any command can be executed. So you can design agent tiers to let the interlocutor understand their current permissions}.” 

Collectively, participants agreed that no single configuration of autonomy is universally appropriate in all teams and organizations. Instead, Manager Clone Agent design must accommodate contextual variation across organizations, teams, and leadership styles. Just as workplace leadership itself is heterogeneous, Manager Clone Agents should preserve space for managers to calibrate their use in ways that best fit their teams. Importantly, dimensions of autonomy do not operate in isolation but in combination. For example, \textit{“a highly realistic Manager Clone Agent that only listens or relays pre-approved messages may still be experienced as low-autonomy”} (M17), while \textit{“a less anthropomorphic text-based agent that intervenes proactively may feel disproportionately intrusive”} (W15). Fine-tuning these combinations is therefore essential for aligning Manager Clone Agent design with individual comfort and organizational needs, while supporting gradual trust-building and long-term adoption.

\subsection{Toward Disruptive and Unsettling Alternative Futures in Speculative Imagination} \label{sec:alternative_future}

While most participant imaginations extended familiar managerial functions, some imagined a set of destabilizing, speculative futures that challenged participants’ own assumptions about hierarchy, authority, and the necessity of human managers altogether. 

\subsubsection{What if Workers Say “No” to Clone Agents?}

Most participants expressed reluctance to use Manager Clone Agents either entirely or in specific situations, yet they also admitted their ability to resist is structurally weak.

Both managers such as M2 and workers such as W1 suggested that open refusal would likely lead to replacement. They agreed on a future where “\textit{Those who say ‘no’ will be filtered out and replaced by workers more willing to comply.}” W14 added: “\textit{It all depends on who has negotiating power. In industries where talent is scarce, workers will successfully refuse.}”

Despite this pessimism, workers articulated concrete conditions for workplace fairness, including access to training data, a third party supported appeals mechanism, veto rights, etc. W3 emphasized: “\textit{The specific reasons behind AI decisions need to be transparent, and we should be informed about what data will be collected about us.}” On the other hand, W15 stresses the need for human managers to be involved in some cases: “\textit{Decisions involving firing, recruiting, or promoting must be reviewed by a human, as they directly affect people’s livelihoods.}” W12 added that a mechanism could be set up to appeal AI decisions, for example, a third party or a labor union could organize collective appeals against decisions. W14 took this speculative future further: “\textit{Such veto power would inevitably run into workplace politics that varies between companies. Workers would try to manipulate any veto power they have, and managers would try to use it to exert as much control as they can. Budgets and roles would be questioned, and confusion would come hand in hand with the speed the clone agents bring.}” 

\subsubsection{What If Managers Fade?}

Several managers including M2, M4, and M5 articulated anxieties that outlined a provocative scenario: a future in which the presence of real managers becomes optional.

This anxiety was captured most vividly in M2’s fiction, which began as a playful thought experiment but drifted, in his words, into “\textit{an apocalyptic film}.” He sadly imaged, “\textit{My clone agent knows their tasks, sometimes better than I do. When I walk by, no one even looks up. They already got what they needed}.” Similarly, M4 feared a future where her Manager Clone Agent “\textit{gets along with the team better than I do,}” absorbing the subtle interpersonal knowledge that she used to define her authority.

The speculative futures raised by managers like M2 and M4 hinted at a more profound philosophical question that cannot be resolved through merely boundary-setting. In M4's views, if clone agents begin to take over the emotional scaffolding, then the anxiety participants described is not about technological encroachment but a deeper ontological shock: what if leadership identity no longer takes “\textit{the unique presence of a person}” as a necessary condition, but become an outsource-able “\textit{set of functions}.” Under this view, M4 believed the challenge is not simply to design against replacement; it is to articulate a new philosophy of management that survives the possibility of replaceability. In this context, it was striking that themes of care, human-centeredness, and relational ethics surfaced repeatedly among managerial participants, as anchors they believed distinguish human leadership from any conceivable clone. As M5 put it, “\textit{If the staff only need instructions, then yes, the agent can do that. But if they need to feel understood as people, they still needs me}.”

Yet even as participants defended these relational ethics as uniquely human, their speculative futures revealed a more unsettling and liberating possibility: What if organizations themselves did not need managers to flourish? As W7 hinted, a workplace mediated by clone agents might buffer harmful interactions, reduce emotional volatility, and allow workers to focus on meaningful work. W3 frankly admitted: “\textit{If the inclusion of AI can diminish, or even overturn power structures like leader–worker dynamics, then as a worker I’m willing to support anything that can facilitate this transformation.}” In such visions, the absence of managers is not a loss but a condition for greater happiness, with more room for individuals to exercise agency.

\subsubsection{What If Clone Agents Become Tools of Resistance?}

Although most workshop discussions centered on how Manager Clone Agents might extend managerial authority, several participants sketched futures in which clone agents are repurposed by workers as instruments of resistance, or even counter-control. These imaginaries surfaced not as fully developed narratives but as hesitant comments and defensive humor.

Workers such as W7 were among the first to hint at this inversion. She desires to use the clone agent as an buffer, shielding herself from emotionally charged interactions. As she put it, “\textit{It can help me keep the evidence straight. If the boss says different things on different days, the agent will know.}” In many other stories participants envisioned, workers use this upward surveillance to advocate for their own interests. In M10’s story, the team frequently accessing the manager’s agent without notice during the manager’s absence, exposing a series of mistakes that ultimately caused the manager to be demoted. M5 jokingly mirrored this risk: “\textit{If my clone keeps a memory of everything I’ve said, it turns into a reflection of me, letting people search for anything they like.}” For some workers such as W3, this was not a joke but a glimpse of a more equitable future. “It makes management and evaluation method transparent.” as W3 commented.

A particularly radical scenario emerged in speculative remarks about “\textit{turning the system around.}” M14 imagined a story where he exploited system transparency. He wrote: “\textit{I’ve figured out what Georgebot’s goals are and I can game the system to get my team more funds.}” After a year, the worker in his story had move on to another company, because of tech turnover, “\textit{I’ll get to say I was trusted with a sizeable budget at a major AI-driven company. Recruiters will see that I worked with AI and they will be impressed. I will continue my embezzlement at my new company, which is also using AI bosses.”}

Across these speculative threads, participants believed if Manager Clone Agents become integral to workplace communication and coordination, workers will inevitably develop ways to appropriate, reinterpret, and strategically mobilize these systems.

\section{Discussion}
This study examined how managers and workers envisioned the possibilities and risks of Manager Clone Agents in future workplaces. Drawing on workshop stories and discussions, we found that participants situated Manager Clone Agents at the intersection of productivity and relations: managers saw them as extensions of their authority and capacity, while workers emphasized how these systems would reshape recognition, fairness, and everyday labor. Across individual, interpersonal, and organizational levels, participants oscillated between hopes of enhanced productivity and fears of displacement or relationships erosion. Building on these findings, we contribute to ongoing debates in HCI and organizational research by situating Manager Clone Agents in three broader conversations: the distinctiveness of clone agents from other algorithmic management systems (see \autoref{boss}), the rethinking of managerial work in the face of automation (see \autoref{work}), and the reconfiguration of worker–manager relations in an automated workplace (see \autoref{relation}). We then discussed the role of design fiction in exploring the future of work and future AI design (see \autoref{fiction}). We also offered design and policy implications towards fair, transparent and accountable AI intermediation (see \autoref{design}).

\subsection{Why Manager Clone Agents Are Not Algorithmic Bosses: A New Mode of AI Authority}\label{boss}

We define Manager Clone Agents as agentic AI systems designed to embody a specific human manager, including their communication style, tone, heuristics, and even symbolic presence~\cite{Liang2025AIClones,lee_speculating_2023,cheng_conversational_2025,journey2025aiAgentsExecutiveProductivity}. Rather than acting as generic assistants, clone agents speak and perform managerial functions “as the manager,” stepping into meetings, clarifying decisions, or guiding workflows in the manager’s absence. Much of the HCI and organizational literature on algorithmic management focuses on algorithmic bosses—systems that assign tasks, monitor workers, and enforce compliance across platform labor and other digitally mediated work settings \cite{mohlmann2021algorithmic,jarrahi2020platformic,lee2016algorithmic}. These systems govern impersonally, relying on metrics, dashboards, and automated procedures. Within this framing, legitimacy is procedural: workers evaluate fairness, transparency, and predictability rather than who the system represents \cite{lee2019procedural,prassl2019if}. By abstracting management away from individual managers, this literature implicitly treats authority as something that can be cleanly encoded in rules and extracted from social relationships. However, such abstractions overlook long-term forms of human authority that derive not from rules but from the social identities and interpersonal histories of specific managers; a gap that Manager Clone Agents make newly visible and bring into AI authority.

Firstly, the Manager Clone Agents re-intermediating algorithmic authority from the procedural domain into the relational and symbolic domain, where legitimacy hinges on social meaning rather than rule adherence. Participants did not judge Manager Clone Agents as faceless systems but as personalized proxies that stand in for a named leader. What mattered was whether it felt like an authentic extension of a specific manager. In our findings, AI-enabled management can be evaluated through relational criteria, for instance, whether the system carries the “voice,” moral sensibility, or interpersonal orientation of the manager. Clone agents expose an important limitation in existing theories of algorithmic management, which largely assume that algorithmic authority operates outside of social identity. By embodying the presence of a specific human, clone agents collapsed the conceptual boundary between algorithmic and human authority.

Secondly, this personalization introduced new vulnerabilities. The same symbolic closeness that granted clone agents legitimacy also raised the stakes when the agent deviated from expected behavior. Participants worried about moments when the agent generated responses the real manager would never give, or adopted tones and stances that felt incongruent. These deviations threatened the coherence of the manager’s identity, leading workers to question whether the AI was still “speaking as” the manager. As shown in previous literature about relational authority \cite{tyler1992relational,wellman2017authority,lake2009relational}, when authority is relational, even small misalignments can destabilize trust in ways procedural errors do not.

Meanwhile, Clone agents also expanded managerial presence in new ways with AI agentic capability. Participants imagined the system representing managerial intent across time zones, onboarding new hires, mediating conflict, or explaining decisions based on inferred managerial logic. In effect, the agent made managerial identity computationally portable. Unlike algorithmic boss systems that impose abstract, depersonalized rules, these agentic capabilities computationally amplified the manager’s own authority and presence, extending what the human manager could accomplish.

Therefore our findings suggest that Manager Clone Agents occupy a conceptual space that existing models of algorithmic bosses do not capture. Legitimacy and trust are therefore tied to the agent’s ability to sustain the social meaning of managerial authority—not merely its ability to optimize workflows or enforce fairness. This reframing challenges HCI and CSCW scholars to expand theories of algorithmic management to account for systems that operate not as neutral infrastructures but as carriers of symbolic power, interpersonal history, and organizational meaning.

\subsection{How Managerial Work and Its Legitimacy Are Reconfigured by Manager Clone Agents}\label{work}

Building on the previous section’s argument that Manager Clone Agents reshape the basis of managerial authority by shifting it from procedural to relational and symbolic grounds, this section turns to what these dynamics mean for the nature and future of managerial work itself. Rather than asking only whether clone agents can legitimately stand in for managers, our findings push us to reconsider a deeper question: what aspects of managerial labor give management its meaning, and how might automation reshape the legitimacy and identity of managerial work? By examining how participants interpreted delegation, amplification, and relational boundaries, we outline three conceptual shifts that challenge prevailing assumptions about what parts of management can or should be automated.

Firstly, our findings complicate the widely assumed distinction between routine managerial tasks suitable for automation and higher-level strategic or relational work that should remain human-led \cite{shao2025future,zaidi2025will,west2018future}. Prior accounts imagine automation as lifting administrative burdens so managers can focus on more substantive leadership \cite{raisch2021artificial}. Yet participants repeatedly emphasized that even tasks that appear routine—such as onboarding a newcomer, acknowledging an effort, or sending a quick clarification—carry symbolic and relational significance. When performed by Manager Clone Agents, these actions can be reinterpreted as a withdrawal of personal engagement, thereby weakening the interpersonal recognition that workers rely on to evaluate managerial legitimacy. What appears functionally automatable may nonetheless remain relationally non-substitutable, revealing that managerial work cannot be cleanly partitioned into “automatable” and “non-automatable” categories without undermining its social meaning.

Secondly, although participants such as W7, M9, and M18 viewed Manager Clone Agents as promising amplifiers of managerial capacity for managers—as discussed in \autoref{sec:amplifier}—they emphasized that this amplification only maintained legitimacy when it respected relational boundaries. Clone agents were accepted when they handled repetitive approvals or relayed standardized updates, enabling managers to redirect their attention toward long-term vision or external relationships. Yet delegating too far into relational or affective domains was described as hollowing out the identity of leadership. This highlights that managerial work is fundamentally relational \cite{sluss2007relational}, and managers must actively choose which interactions to keep as human-only encounters, knowing that these choices directly shape their perceived legitimacy and the integrity of their leadership identity.

Thirdly, our findings point toward a more speculative but increasingly plausible future in which the experience of being managed becomes a hybrid interactional ecology jointly mediated by humans and AI. Across workshops, participants imagined scenarios where workers interacted first, or even predominantly, with the clone agent rather than the human manager. If such patterns become normalized, managerial authority may depend less on managers’ in-person relational performances and more on their stewardship of the digital systems that now partially represent them. In such a future, managers are judged not only by how they lead directly, but by how responsibly they configure the tone, boundaries, and comportment of their AI counterparts.  Legitimacy thus becomes distributed across the manager, the system’s reliability, and the manager’s governance of that system. In this context, managerial legitimacy shifts toward an infrastructural form: rooted in how managers design, maintain, and ethically govern the socio-technical environments through which workers experience leadership.

\subsection{How AI Managerial Agents Reshape Labor, Power, and Worker Agency}   \label{relation} 

Building on our earlier analysis of managerial authority, this section turns to the worker–manager relationship and asks how Manager Clone Agents reshape labor, power, and worker agency in AI-mediated organizations. Like in other labor studies literature \cite{barati2022effects,curchod2020working,monod2025robots}, the relationship between managers and workers in our study is structured by deep and persistent power asymmetries in authority, evaluative control, access to information, and opportunities for visibility. Our findings consistently revealed these asymmetries in practice. Managers enjoyed privileged channels of recognition and decision-making: they delegated tasks, set priorities, interpreted performance, and decided when to appear or withdraw from everyday interactions. Workers, by contrast, relied on scarce moments of face-time to demonstrate competence, depended on managers to relay their concerns upward, and navigated the risks of being misunderstood, overlooked, or evaluated unfairly. Within this already asymmetric terrain, Manager Clone Agents do not arrive into a neutral field—they enter a workplace where one side holds disproportionate control over access, evaluation, and agenda-setting. Existing power inequalities still shape how clone agents are deployed, and how their effects are experienced.

However, parallel strands of research highlight that workers are not passive recipients of technological control. Workers actively develop tactics of resistance, appropriation, and reinterpretation in response to algorithmic systems \cite{ettlinger2018algorithmic,kellogg2020algorithms,heemsbergen2022introduction}. Our findings also reveal a far more dynamic and contested terrain. Rather than a simple story of top-down authority, Manager Clone Agents complicate the relational fabric of work by redistributing visibility, recognition, and informational power in uneven and sometimes unexpected ways. To illuminate these tensions, we outline three shifts in how workers positively imagine their relationship to managers in a clone-mediated workplace.

First, Manager Clone Agents reshape the baseline power relation between workers and managers by redistributing how authority, access, and recognition circulate in everyday work. While managers in our study largely embraced clone agents as extensions of their authority and capacity, workers evaluated these same systems through a very different set of concerns. They focused on whether clone agents preserved meaningful access to managers, ensured fair visibility for their contributions, and supported rather than undermined their career opportunities.

Second, beyond these structural shifts, workers imagined immediate, everyday forms of agency emerging through their interactions with clone agents. They envisioned using agents to clarify ambiguous expectations, document contradictory instructions, and buffer emotionally difficult interactions. In this sense, clone agents become relational intermediaries that workers can strategically mobilize to protect themselves, secure clarity, or ensure commitments are honored. Such micro-practices illustrate how worker agency persists, and even expands, within AI-mediated management through subtle, everyday acts of reinterpretation and tactical use.

Third, participants articulated more radical future-oriented forms of counter-control, imagining scenarios where clone agents become instruments for challenging managerial opacity or renegotiating organizational power relations. In these speculative futures, workers leverage the agent’s persistent memory, consistency, and legibility to audit managerial behavior, hold managers accountable, or strategically navigate decision-making systems. Some imagined using clone agents to expose biases or inconsistencies; others envisioned querying the agent for rationales behind past decisions, thereby transforming managerial authority into something searchable and contestable. More radical imaginaries involved workers gaming the system to secure resources or reshape how managerial performance is evaluated. These visions extend everyday resistance into a broader reconfiguration of organizational power, suggesting that clone agents may enable workers to practice new forms of counter-control rooted in upward transparency and data-driven accountability \cite{kristiansen2018accountability,sun2023care}.

\subsection{The Role of Design Fiction: Speculating the Agent’s Place in Future Work} \label{fiction}

In this study, design fiction served as more than a vehicle for imagining future technologies; it operated as a method for making future organizational possibilities thinkable \cite{wong2021using,ringfort2023design}. By introducing Manager Clone Agents into speculative workplace scenarios, design fiction enabled participants to articulate relational and power-based dynamics that are difficult to surface through traditional interview or observational methods.

Firstly, the fictional frame created a form of temporal and psychological distance that allowed participants to bracket present-day organizational constraints \cite{sloan2025temporal}. This distance enabled them to articulate forms of authority, recognition, and legitimacy that are normally taken for granted in everyday work. By stepping outside current norms, participants could ask questions that are not yet empirically available, such as how a computationally embodied manager might affect identity, trust, or the felt presence of leadership.

Secondly, design fiction opened a space for divergent and even contradictory future imaginaries \cite{tanenbaum2016limits}. Rather than extending current algorithmic management trends, the speculative format elicited futures that ranged from incremental modifications to radically reorganized hierarchies. This plurality is essential: by engaging with futures that challenge or invert existing power structures, participants were able to articulate what must be protected (e.g., relational boundaries), what must be supported (e.g., worker agency), and what must be constrained (e.g., over-delegation of relational labor). Thus, design fiction not only surfaces multiple futures but also clarifies how managerial agents should be designed to remain accountable, relationally appropriate, and equitable across those futures.

Thirdly, design fiction functioned as an analytic probe \cite{schulte2016homes}. The stories participants generated were not treated as predictions but as structured reflections that expose the relational, ethical, and organizational stakes embedded in emerging managerial agents. Through these narrative probes, we were able to observe how people reason about delegation limits, how they interpret the boundaries of managerial presence, and how they imagine exercising or retaining agency long before such agents materialize in practice.

\subsection{Design and Policy Implications: Towards Fair, Transparent and Accountable AI Intermediation}  \label{design}

Beyond theoretical contributions, our findings also highlight several implications for the design and governance of Manager Clone Agents. From a design perspective, Manager Clone Agents should incorporate mechanisms that clarify boundaries and preserve human agency. Participants valued augmentation but resisted substitution, suggesting that systems should make it explicit when an interaction is mediated by an agent, what kinds of contributions the agent is authorized to make, and how accountability is distributed. This requires visible and persistent disclosure cues, such as labeled message headers, distinctive visual avatars, or interaction summaries, that help workers distinguish when they are engaging with a clone versus the human manager. Our findings further indicate that workers evaluate legitimacy based on the perceived “domain” of the interaction; thus, interfaces should allow managers to designate certain contexts (e.g., onboarding, conflict navigation, performance feedback) as human-only zones and automatically prevent agent participation in those settings.

Design strategies could also include configurable tiers of initiative, allowing managers to set limits on whether the agent may proactively answer questions, surface insights, or merely relay prepared statements. For example, a clone agent might be restricted to providing clarifications in low-stakes administrative queries but require explicit manager approval before making recommendations in ambiguous or relationally sensitive situations. Workers in our study also emphasized the need for recourse: interfaces should provide quick mechanisms for requesting escalation to the human manager, annotating moments of misunderstanding, or pausing agent involvement when relational trust is at risk. Additionally, because many workers used clone agents as tools for emotional buffering, design should support private annotation spaces where workers can rehearse drafts, explore emotionally charged questions, or reflect before escalating to a human.

From a policy perspective, organizations adopting Manager Clone Agents must create safeguards that protect workers from new vulnerabilities. This includes ensuring that workers’ reliance on or resistance to Manager Clone Agents does not become a hidden basis for performance evaluation, and that managerial delegation does not erode opportunities for worker recognition and advancement. Policies should mandate enforceable worker controls: workers must be able to configure binding preferences that prevent agent participation in designated contexts (e.g., evaluation, promotion) without requiring managerial approval for each instance. Organizational guidelines must also require managers to retain responsibility for decisions made in their name, prohibiting the use of clone agents as shields for deflecting blame or obscuring authorship.

Transparency policies should mandate clear documentation of when clone agents are active, what training data they draw from, and what logs are retained, especially because workers in our study used agent memory to track fairness or resolve inconsistencies. Organizations should provide workers with controlled access to interaction histories involving clone agents so they can contest errors, ensure accurate representation, and prevent managerial misuse of selective memory.

At a broader level, regulatory frameworks may need to address the dual identity of Manager Clone Agents as both tools and proxies, clarifying where legal and ethical responsibility lies when agents act in a manager’s name. This could include requirements for audit trails that preserve the boundary between human intent and agent action, guidelines for preventing over-delegation of relational labor, and protections against reproducing or amplifying managerial biases in cloned decision patterns. Ultimately, governance must ensure that Manager Clone Agents enhance rather than erode relational equity: workers must have concrete mechanisms for contestation: the ability to flag and appeal agent decisions through channels that do not depend on managerial discretion, access to agent decision rationales, and formal representation in governance decisions about agent deployment.

\section{Limitations and Future Work}  

Like any qualitative and speculative study, our work has several limitations that should be acknowledged. First, our findings are based on six design fiction workshops with 23 participants. While this approach enabled us to surface rich imaginaries and relational concerns, the sample is not representative of all industries, organizational sizes, or cultural contexts. Second, our use of speculative design elicited visions of possibilities and risks rather than documenting actual deployments. This allowed participants to reflect openly without being constrained by current technological limits, but it also means that their narratives may differ from how Manager Clone Agents are eventually integrated in practice. Third, our focus was primarily on managers and workers situated within formal organizations. Other stakeholders, such as HR professionals, policymakers, or AI developers, are likely to shape how Manager Clone Agents are designed and governed. Expanding the scope to include these perspectives would enrich understanding of the broader ecosystem in which Manager Clone Agents operate. Fourth, our study's scope limits our ability to fully operationalize these principles into concrete artifact-level mechanisms without deeper engagement with workers' lived experiences of power asymmetries. Similarly, the difficulty participants faced in imagining structural alternatives to existing hierarchies within mixed manager-worker workshops is notable. Future work could explicitly target these gaps through workers-only speculative sessions and participatory design focused on the practical mechanics of contestation and control.

\section{Conclusion}  

This paper examined how managers and workers envisioned the opportunities and risks of Manager Clone Agents, AI-powered surrogates trained on managerial communication and practices to act on behalf of leaders. Through six design fiction workshops with 23 participants, we identified potential supportive roles in future workplaces and the tensions they created across individual, interpersonal, and organizational levels. We argue that the design and governance of Manager Clone Agents must be grounded in relational awareness, ensuring that these systems enhance rather than replace the symbolic, affective, and interpersonal practices that give managerial work meaning.

\bibliographystyle{ACM-Reference-Format}
\bibliography{zotero,bibliography}

@incollection{wong2018speculative,
  title={Speculative design in HCI: from corporate imaginations to critical orientations},
  author={Wong, Richmond Y and Khovanskaya, Vera},
  booktitle={New Directions in Third Wave Human-Computer Interaction: Volume 2-Methodologies},
  pages={175--202},
  year={2018},
  publisher={Springer}
}

@book{dunne2024speculative,
  title={Speculative Everything, With a new preface by the authors: Design, Fiction, and Social Dreaming},
  author={Dunne, Anthony and Raby, Fiona},
  year={2024},
  publisher={MIT press}
}

@inproceedings{ma2025speculative,
  title={Speculative Job Design: Probing Alternative Opportunities for Gig Workers in an Automated Future},
  author={Ma, Shuhao and Liu, Zhiming and Nisi, Valentina and Fox, Sarah E and Nunes, Nuno Jardim},
  booktitle={Proceedings of the 2025 CHI Conference on Human Factors in Computing Systems},
  pages={1--18},
  year={2025}
}

@article{Gani2025ai_hotline,
  author       = {Aisha S. Gani},
  title        = {When customers dial Klarna’s hotline, an AI CEO picks up},
  journal      = {Bloomberg News (newsletter)},
  year         = {2025},
  month        = {September},
  day          = {10},
}

@article{journey2025aiAgentsExecutiveProductivity,
  author       = {{The Journey Platform}},
  title        = {How AI agents are redefining executive productivity},
  journal      = {The Journey Platform},
  year         = {2025},
  month        = {Apr},
  note         = {Accessed: 2025-09-07},
  url          = {https://thejourneyplatform.com/blog-posts/how-ai-agents-are-redefining-executive-productivity}
}

@article{forsdick2025aiavatars,
  author       = {Sam Forsdick},
  title        = {Tech CEOs are sending their AI avatars to meetings},
  journal      = {Raconteur},
  year         = {2025},
  month        = may,
  day          = {29},
  url          = {https://www.raconteur.net/technology/ai-avatars-meetings},
  note         = {Accessed: 2025-09-07}
}

@article{synthesia2025seriesD,
  author       = {Lunden, Ingrid},
  title        = {Synthesia Snaps Up \$180 Million at a \$2.1 B Valuation for Its B2B AI Video Platform},
  journal      = {TechCrunch},
  date         = {2025-01-14},
  note         = {Funding round led by NEA, with participation from WiL, Atlassian Ventures, PSP Growth, plus GV and MMC Ventures; total raised to date \$330 million; 60,000 businesses, 1 million users}  
}

@article{elevenlabs2025seriesC,
  author       = {Mehta, Ivan and Lunden, Ingrid},
  title        = {ElevenLabs, the Hot AI Audio Startup, Confirms \$180 Million in Series C Funding at a \$3.3 B Valuation},
  journal      = {TechCrunch},
  date         = {2025-01-30},
  note         = {co-led by a16z and ICONIQ Growth, total raised to date \$281 million}  
}

@misc{delphi2025seriesA,
  author       = {Ladjevardian, Dara},
  title        = {Delphi Raises \$16M Series A from Sequoia Capital to Pioneer “Digital Minds”},
  howpublished = {Blog post, Delphi AI},
  year         = {2025},
  month        = jun # "~24",
  url          = {https://www.delphi.ai/blog/delphi-raises-16m-series-a-from-sequoia}
}

@misc{heygen2025seriesA,
  author       = {HeyGen Team},
  title        = {HeyGen Secures \$60 M Series A to Power AI Video Generation for Business Growth},
  howpublished = {Blog post},
  institution  = {HeyGen},
  date         = {2025-08-19},
  note         = {led by Benchmark, includes investors Thrive Capital, BOND, SV Angel; company jumped from \$1 M to \$35 M+ ARR}  
}

@article{Bonos_Abril_2025, 
  title   = {No one likes meetings. They’re sending their AI note takers instead.}, 
  author  = {Bonos, Lisa and Abril, Danielle}, 
  journal = {The Washington Post}, 
  year    = {2025}, 
  url     = {https://www.washingtonpost.com/technology/2025/07/02/ai-note-takers-meetings-bots/}
}

@online{maslworld2025middlemanager,
  title         = {AI Is Replacing the Middle Manager—Quietly and Relentlessly [2025 Workforce Shift]},
  author        = {{MASL WORLD}},
  year          = {2025},
  date          = {2025-04-19},
  organization  = {LinkedIn},
  subtitle      = {MASL-NEWSLETTER TECHBRIDGE},
  url           = {https://www.linkedin.com/pulse/ai-replacing-middle-manager-quietly-relentlessly-2025-workforce-lpg0e/},
  note          = {Accessed: 2025-09-04}
}

@article{raconteur2023ai,
  title={Will AI transform the CEO’s role or replace it entirely?},
  author={{Raconteur}},
  journal={Raconteur},
  year={2023},
  url={https://www.raconteur.net/technology/ai-transform-replace-ceo-role}
}

@misc{nvidia2022omniverse,
  author       = {NVIDIA},
  title        = {NVIDIA Omniverse Avatar Platform for Building Real Time, Interactive, AI Assistants},
  year         = {2022},
  howpublished = {\url{https://www.youtube.com/watch?v=ETWTbmi0W20}},
  note         = {Accessed: 2025-08-30},
}

@online{heygen-reidai,
  title = {Reid Hoffman’s Digital Twin: How HeyGen Powers AI-Driven Presence},
  author = {{HeyGen}},
  year = {2024},
  url = {https://www.heygen.com/customer-stories/reid-ai},
  note = {Accessed: 2025-08-30},
}

@article{patel_zoom_ceo_2024,
  title   = {Zoom CEO Eric Yuan wants AI clones in meetings},
  author  = {Nilay Patel},
  journal = {The Verge},
  year    = {2024},
  month = jun,
  url     = {https://www.theverge.com/2024/6/3/24168733/zoom-ceo-ai-clones-digital-twins-videoconferencing-decoder-interview},
  note    = {Accessed: 2025-08-30},
}

@online{morrone2025ai,
  author    = {Megan Morrone},
  title     = {CEOs using AI avatars and digital clones as workers fear job loss from AI},
  year      = {2025},
  date      = {2025-06-10},
  url       = {https://www.axios.com/2025/06/10/ai-ceo-clones-trust-gap},
  note      = {Axios},
}

@online{bort2025klarna,
  author    = {Julie Bort},
  title     = {Klarna used an AI avatar of its CEO to deliver earnings, it said},
  year      = {2025},
  url       = {https://techcrunch.com/2025/05/21/klarna-used-an-ai-avatar-of-its-ceo-to-deliver-earnings-it-said/},
  note      = {Accessed: 2025-08-30},
  organization = {TechCrunch}
}

@online{peters2025techceos,
  author       = {Jay Peters},
  title        = {Tech CEOs are using AI to replace themselves},
  year         = {2025},
  month        = {May},
  url          = {https://www.theverge.com/news/673194/tech-ceos-zoom-klarna-replace-earnings},
  organization = {The Verge},
  note         = {Accessed: 2025-08-30}
}

@article{boyle2025aiCEO,
  author    = {Matthew Boyle},
  title     = {Meetings Won’t Be the Same When the CEO Sends an AI Bot},
  journal   = {Bloomberg},
  url       = {https://www.bloomberg.com/news/features/2025-04-15/meetings-won-t-be-the-same-when-the-ceo-sends-an-ai-bot},
  date      = {2025-04-15},
  note      = {Accessed: 2025-08-30}
}

@online{personalai_salomon_2025,
  title = {How Salomon Scaled Its CEO with Personal AI—While Slashing Comms Time by 75\%},
  author = {{Personal AI}},
  year = {2025},
  url = {https://www.personal.ai/customers/salomon},
  note = {Accessed: 2025-08-30},
}

@online{Langley2025Lattice,
  author       = {Hugh Langley},
  title        = {How Lattice Is Preparing for Humans and AI Agents to Work Together},
  year         = {2025},
  month        = {June},
  url          = {https://www.businessinsider.com/lattice-ai-agents-humans-work-jobs-2025-6},
  organization = {Business Insider},
  note         = {Accessed: 2025-08-30}
}

@online{cuthbertson2023company,
  author    = {Anthony Cuthbertson},
  title     = {Company that made an AI its chief executive sees stocks climb},
  year      = {2023},
  date      = {2023-03-16},
  url       = {https://www.the-independent.com/tech/ai-ceo-artificial-intelligence-b2302091.html},
  note      = {The Independent}
}

@techreport{pwcAIAgentSurvey2025,
  title        = {AI Agent Survey: How U.S. companies are deploying value-generating AI agents},
  institution  = {PricewaterhouseCoopers (PwC) US},
  author    = {{PwC}},
  year         = {2025},
  url          = {https://www.pwc.com/us/en/tech-effect/ai-analytics/ai-agent-survey.html},
}

@misc{sequoia2025delphi,
  title = {How Delphi’s AI Digital Minds Can Scale Human Connection},
  author = {Ladjevardian, Dara and Lee, Jess and Huang, Sonya},
  year = {2025},
  note = {Podcast episode 58, Training Data, Sequoia Capital. Available at: \url{https://www.sequoiacap.com/podcast/training-data-dara-ladjevardian/}},
}

@article{FiferFeagans2025,
  title = {Lower AI Costs Will Drive Innovation, Efficiency, and Adoption},
  author = {Bo Fifer and Evan W. Feagans},
  journal = {TCW Insights},
  year = {2025},
  month = {February},
  url = {https://www.tcw.com/Insights/2025/2025-02-19-Thematic-AI}
}

@online{digitaldefynd2025ai,
  author    = {Team DigitalDefynd},
  title     = {Impact of AI on Middle Management [2025]},
  year      = {2025},
  url       = {https://digitaldefynd.com/IQ/ai-in-middle-management/},
  note      = {Accessed: 2025-08-30}
}

@article{Sowa_Przegalinska_Ciechanowski_2020, title={Cobots in knowledge work}, volume={125}, url={https://www.sciencedirect.com/science/article/pii/S014829632030792X?via%3Dihub}, DOI={10.1016/j.jbusres.2020.11.038}, journal={Journal of Business Research}, author={Sowa, Konrad and Przegalinska, Aleksandra and Ciechanowski, Leon}, year={2020}, month=dec, pages={135–142} }

@techreport{writer2025generativeai,
  author    = {Writer},
  title     = {Generative AI Adoption in the Enterprise},
  institution = {Writer},
  year      = {2025},
  type      = {Industry Report},
  url       = {https://go.writer.com/hubfs/pdfs/generative-ai-adoption-enterprise-survey-writer-com.pdf?hsLang=en}
}

@book{Mintzberg_1973, title={The nature of managerial work}, url={http://hib510week9.pbworks.com/f/The+Nature+of+Managerial+Work,+Mintzberg+1973.pdf}, publisher={Harper \& Row}, author={Mintzberg, H.}, year={1973} }

@article{dler2021importance,
  title={Importance of Managerial Roles and Capabilities on Organizational Effectiveness},
  author={Dler, Shekh Mohammed and Tawfeq, Akam Omar},
  journal={International Journal of Academic Research in Business and Social Sciences},
  volume={11},
  number={4},
  pages={957--964},
  year={2021}
}

@article{stray2020understanding,
  title={Understanding coordination in global software engineering: A mixed-methods study on the use of meetings and Slack},
  author={Stray, Viktoria and Moe, Nils Brede},
  journal={Journal of Systems and Software},
  volume={170},
  pages={110717},
  year={2020},
  publisher={Elsevier}
}

@article{Andersone_Nardelli_Ipsen_Edwards_2022, title={Exploring Managerial job demands and Resources in Transition to distance Management: a qualitative Danish case study}, volume={20}, url={https://www.mdpi.com/1660-4601/20/1/69}, DOI={10.3390/ijerph20010069}, number={1}, journal={International Journal of Environmental Research and Public Health}, author={Andersone, Nelda and Nardelli, Giulia and Ipsen, Christine and Edwards, Kasper}, year={2022}, month=dec, pages={69} }

@article{morrison2020challenges,
  title={Challenges and barriers in virtual teams: a literature review},
  author={Morrison-Smith, Sarah and Ruiz, Jaime},
  journal={SN Applied Sciences},
  volume={2},
  number={6},
  pages={1096},
  year={2020},
  publisher={Springer}
}

@article{choudhury2023virtual,
  title={Virtual water coolers: A field experiment on the role of virtual interactions on organizational newcomer performance},
  author={Choudhury, Prithwiraj and N Lane, Jacqueline and Bojinov, Iavor},
  journal={Harvard Business School Technology \& Operations Mgt. Unit Working Paper},
  number={21-125},
  year={2023}
}

@inproceedings{grudin1988cscw,
  title={Why CSCW applications fail: problems in the design and evaluationof organizational interfaces},
  author={Grudin, Jonathan},
  booktitle={Proceedings of the 1988 ACM conference on Computer-supported cooperative work},
  pages={85--93},
  year={1988}
}

@incollection{YuklVanFleet1992,
  author    = {Gary Yukl and David D. Van Fleet},
  title     = {Theory and Research on Leadership in Organizations},
  booktitle = {Handbook of Industrial and Organizational Psychology},
  editor    = {Marvin D. Dunnette and Leaetta M. Hough},
  publisher = {Consulting Psychologists Press},
  year      = {1992},
  pages     = {147--197},
  url       = {https://www.researchgate.net/publication/286930761_Theory_and_Research_on_Leadership_in_Organizations},
}

@book{tamkin, title={Crowdsourced leadership IES Perspectives on HR 2014}, url={https://www.employment-studies.co.uk/system/files/resources/files/mp102.pdf}, author={Tamkin, Penny}, publisher={Institute for Employment Studies}, year={2014},
}

@article{meagher2020worker,
  title={Worker trust in management and delegation in organizations},
  author={Meagher, Kieron J and Wait, Andrew},
  journal={The Journal of Law, Economics, and Organization},
  volume={36},
  number={3},
  pages={495--536},
  year={2020},
  publisher={Oxford University Press}
}

@article{vuori2025s,
  title={It's Amazing--But Terrifying!: Unveiling the Combined Effect of Emotional and Cognitive Trust on Organizational Member'Behaviours, AI Performance, and Adoption},
  author={Vuori, Natalia and Burkhard, Barbara and Pitk{\"a}ranta, Leena},
  journal={Journal of Management Studies},
  year={2025},
  publisher={Wiley Online Library}
}

@article{cheon2022working,
  title={Working with bounded collaboration: A qualitative study on how collaboration is co-constructed around collaborative robots in industry},
  author={Cheon, EunJeong and Schneiders, Eike and Skov, Mikael B},
  journal={Proceedings of the ACM on Human-Computer Interaction},
  volume={6},
  number={CSCW2},
  pages={1--34},
  year={2022},
  publisher={ACM New York, NY, USA}
}

@article{wolf2025one,
  title={One size does not fit all: Mechanisms of employees’ acceptance of robotic lower-level managers},
  author={Wolf, Franziska D and Stock-Homburg, Ruth M},
  journal={Group \& Organization Management},
  pages={10596011251313568},
  year={2025},
  publisher={SAGE Publications Sage CA: Los Angeles, CA}
}

@inproceedings{lee2015working,
  title={Working with machines: The impact of algorithmic and data-driven management on human workers},
  author={Lee, Min Kyung and Kusbit, Daniel and Metsky, Evan and Dabbish, Laura},
  booktitle={Proceedings of the 33rd annual ACM conference on human factors in computing systems},
  pages={1603--1612},
  year={2015}
}

@article{rostron2022hero,
  title={How to be a hero: How managers determine what makes a good manager through narrative identity work},
  author={Rostron, Ali},
  journal={Management Learning},
  volume={53},
  number={3},
  pages={417--438},
  year={2022},
  publisher={Sage Publications Sage UK: London, England}
}

@article{cao2021my,
  title={My team will go on: Differentiating high and low viability teams through team interaction},
  author={Cao, Hancheng and Yang, Vivian and Chen, Victor and Lee, Yu Jin and Stone, Lydia and Diarrassouba, N'godjigui Junior and Whiting, Mark E and Bernstein, Michael S},
  journal={Proceedings of the ACM on Human-Computer Interaction},
  volume={4},
  number={CSCW3},
  pages={1--27},
  year={2021},
  publisher={ACM New York, NY, USA}
}

@inproceedings{10.1145/3613904.3642114,
author = {Li, Jie and Cao, Hancheng and Lin, Laura and Hou, Youyang and Zhu, Ruihao and El Ali, Abdallah},
title = {User Experience Design Professionals’ Perceptions of Generative Artificial Intelligence},
year = {2024},
isbn = {9798400703300},
publisher = {Association for Computing Machinery},
address = {New York, NY, USA},
url = {https://doi.org/10.1145/3613904.3642114},
doi = {10.1145/3613904.3642114},
booktitle = {Proceedings of the 2024 CHI Conference on Human Factors in Computing Systems},
articleno = {381},
numpages = {18},
location = {Honolulu, HI, USA},
series = {CHI '24}
}

@misc{Schmitt_2024, title={Strategic Integration of Artificial Intelligence in the C-Suite: The role of the Chief AI Officer}, url={https://arxiv.org/abs/2407.10247}, journal={arXiv.org}, author={Schmitt, Marc}, year={2024}, month=apr }

@article{lee2018understanding,
  title={Understanding perception of algorithmic decisions: Fairness, trust, and emotion in response to algorithmic management},
  author={Lee, Min Kyung},
  journal={Big data \& society},
  volume={5},
  number={1},
  pages={2053951718756684},
  year={2018},
  publisher={SAGE Publications Sage UK: London, England}
}

@article{rosenblat2016algorithmic,
  title={Algorithmic labor and information asymmetries: A case study of Uber’s drivers},
  author={Rosenblat, Alex and Stark, Luke},
  journal={International journal of communication},
  volume={10},
  pages={27},
  year={2016}
}

@article{park2024generative,
  title={Generative agent simulations of 1,000 people},
  author={Park, Joon Sung and Zou, Carolyn Q and Shaw, Aaron and Hill, Benjamin Mako and Cai, Carrie and Morris, Meredith Ringel and Willer, Robb and Liang, Percy and Bernstein, Michael S},
  journal={arXiv preprint arXiv:2411.10109},
  year={2024}
}

@inproceedings{cranshaw2017calendar,
  title={Calendar. help: Designing a workflow-based scheduling agent with humans in the loop},
  author={Cranshaw, Justin and Elwany, Emad and Newman, Todd and Kocielnik, Rafal and Yu, Bowen and Soni, Sandeep and Teevan, Jaime and Monroy-Hern{\'a}ndez, Andr{\'e}s},
  booktitle={Proceedings of the 2017 CHI Conference on Human Factors in Computing Systems},
  pages={2382--2393},
  year={2017}
}

@article{Liang2025AIClones,
  author    = {Annie Liang},
  title     = {Artificial Intelligence Clones},
  journal   = {arXiv preprint arXiv:2501.16996},
  year      = {2025},
  month     = apr,
}

@article{braun2021one,
  title={One size fits all? What counts as quality practice in (reflexive) thematic analysis?},
  author={Braun, Virginia and Clarke, Victoria},
  journal={Qualitative research in psychology},
  volume={18},
  number={3},
  pages={328--352},
  year={2021},
  publisher={Taylor \& Francis}
}

@inproceedings{fan2025user,
  title={User-Driven Value Alignment: Understanding Users' Perceptions and Strategies for Addressing Biased and Discriminatory Statements in AI Companions},
  author={Fan, Xianzhe and Xiao, Qing and Zhou, Xuhui and Pei, Jiaxin and Sap, Maarten and Lu, Zhicong and Shen, Hong},
  booktitle={Proceedings of the 2025 CHI Conference on Human Factors in Computing Systems},
  pages={1--19},
  year={2025}
}

@inproceedings{xiao2025might,
  title={" It Might be Technically Impressive, But It's Practically Useless to us": Motivations, Practices, Challenges, and Opportunities for Cross-Functional Collaboration around AI within the News Industry},
  author={Xiao, Qing and Fan, Xianzhe and Simon, Felix Marvin and Zhang, Bingbing and Eslami, Motahhare},
  booktitle={Proceedings of the 2025 CHI Conference on Human Factors in Computing Systems},
  pages={1--19},
  year={2025}
}

@article{braun2019reflecting,
  title={Reflecting on reflexive thematic analysis},
  author={Braun, Virginia and Clarke, Victoria},
  journal={Qualitative research in sport, exercise and health},
  volume={11},
  number={4},
  pages={589--597},
  year={2019},
  publisher={Taylor \& Francis}
}

@article{campbell2020purposive,
  title={Purposive sampling: complex or simple? Research case examples},
  author={Campbell, Steve and Greenwood, Melanie and Prior, Sarah and Shearer, Toniele and Walkem, Kerrie and Young, Sarah and Bywaters, Danielle and Walker, Kim},
  journal={Journal of research in Nursing},
  volume={25},
  number={8},
  pages={652--661},
  year={2020},
  publisher={Sage Publications Sage UK: London, England}
}

@article{green2019principles,
  title={The principles and limits of algorithm-in-the-loop decision making},
  author={Green, Ben and Chen, Yiling},
  journal={Proceedings of the ACM on human-computer interaction},
  volume={3},
  number={CSCW},
  pages={1--24},
  year={2019},
  publisher={ACM New York, NY, USA}
}

@inproceedings{zhang2022a,
  author    = {Zhang, Angie and Boltz, Alexander and Wang, Chun Wei and Lee, Min Kyung},
  title     = {Algorithmic Management Reimagined For Workers and By Workers: Centering Worker Well‑Being in Gig Work},
  booktitle = {Proceedings of the 2022 CHI Conference on Human Factors in Computing Systems},
  year      = {2022},
  pages     = {1--20},
}

@article{spektor2023charting,
  author  = {Spektor, Franchesca and Fox, Sarah E. and Awumey, Ezra and Begleiter, Ben and Kulkarni, Chinmay and Stringam, Betsy and Riordan, Christine A. and Rho, Hye Jin and Akridge, Hunter and Forlizzi, Jodi},
  title   = {Charting the Automation of Hospitality: An Interdisciplinary Literature Review Examining the Evolution of Frontline Service Work in the Face of Algorithmic Management},
  journal = {Proceedings of the ACM on Human‑Computer Interaction},
  volume  = {7},
  number  = {CSCW1},
  year    = {2023},
  article = {33},
}

@article{li2025metaagents,
  title={MetaAgents: Large Language Model Based Agents for Decision-Making on Teaming},
  author={Li, Yuan and Sun, Lichao and Zhang, Yixuan},
  journal={Proceedings of the ACM on Human-Computer Interaction},
  volume={9},
  number={2},
  pages={1--27},
  year={2025},
  publisher={ACM New York, NY, USA}
}

@inproceedings{kim2020bot,
  title={Bot in the bunch: Facilitating group chat discussion by improving efficiency and participation with a chatbot},
  author={Kim, Soomin and Eun, Jinsu and Oh, Changhoon and Suh, Bongwon and Lee, Joonhwan},
  booktitle={Proceedings of the 2020 CHI Conference on Human Factors in Computing Systems},
  pages={1--13},
  year={2020}
}

@article{benke2020chatbot,
  title={Chatbot-based emotion management for distributed teams: A participatory design study},
  author={Benke, Ivo and Knierim, Michael Thomas and Maedche, Alexander},
  journal={Proceedings of the ACM on Human-Computer Interaction},
  volume={4},
  number={CSCW2},
  pages={1--30},
  year={2020},
  publisher={ACM New York, NY, USA}
}

@article{do2022should,
  title={How should the agent communicate to the group? Communication strategies of a conversational agent in group chat discussions},
  author={Do, Hyo Jin and Kong, Ha-Kyung and Lee, Jaewook and Bailey, Brian P},
  journal={Proceedings of the ACM on Human-Computer Interaction},
  volume={6},
  number={CSCW2},
  pages={1--23},
  year={2022},
  publisher={ACM New York, NY, USA}
}

@inproceedings{schulte2016homes,
  title={Homes for life: a design fiction probe},
  author={Schulte, Britta F and Marshall, Paul and Cox, Anna L},
  booktitle={Proceedings of the 9th nordic conference on human-computer interaction},
  pages={1--10},
  year={2016}
}

@inproceedings{tanenbaum2016limits,
  title={The limits of our imagination: design fiction as a strategy for engaging with dystopian futures},
  author={Tanenbaum, Theresa Jean and Pufal, Marcel and Tanenbaum, Karen},
  booktitle={Proceedings of the Second Workshop on Computing within Limits},
  pages={1--9},
  year={2016}
}

@inproceedings{wong2021using,
  title={Using design fiction memos to analyze ux professionals’ values work practices: A case study bridging ethnographic and design futuring methods},
  author={Wong, Richmond Y},
  booktitle={Proceedings of the 2021 CHI Conference on Human Factors in Computing Systems},
  pages={1--18},
  year={2021}
}

@article{sloan2025temporal,
  title={Temporal Navigation: How Organizations Work with Multiple Futures},
  author={Sloan, Jennifer N},
  year={2025}
}

@inproceedings{ringfort2023design,
  title={Design Fiction in a Corporate Setting--a Case Study},
  author={Ringfort-Felner, Ronda and Neuhaus, Robin and D{\"o}rrenb{\"a}cher, Judith and Gro{\ss}kopp, Sabrina and Theofanou-F{\"u}lbier, Dimitra and Hassenzahl, Marc},
  booktitle={Proceedings of the 2023 ACM Designing Interactive Systems Conference},
  pages={2093--2108},
  year={2023}
}

@article{monod2025robots,
  title={Robots as Managers? Organizational Justice, Power Asymmetries, and the Future of AI Supervision},
  author={Monod, Emmanuel},
  journal={Management Research Quarterly},
  volume={2},
  number={1},
  pages={41--49},
  year={2025}
}

@article{curchod2020working,
  title={Working for an algorithm: Power asymmetries and agency in online work settings},
  author={Curchod, Corentin and Patriotta, Gerardo and Cohen, Laurie and Neysen, Nicolas},
  journal={Administrative science quarterly},
  volume={65},
  number={3},
  pages={644--676},
  year={2020},
  publisher={Sage Publications Sage CA: Los Angeles, CA}
}

@article{barati2022effects,
  title={Effects of algorithmic control on power asymmetry and inequality within organizations},
  author={Barati, Mehdi and Ansari, Bahareh},
  journal={Journal of Management Control},
  volume={33},
  number={4},
  pages={525--544},
  year={2022},
  publisher={Springer}
}

@article{wellman2017authority,
  title={Authority or community? A relational models theory of group-level leadership emergence},
  author={Wellman, Ned},
  journal={Academy of Management Review},
  volume={42},
  number={4},
  pages={596--617},
  year={2017},
  publisher={Academy of Management Briarcliff Manor, NY}
}

@article{sun2023care,
  title={Care Workers' Wellbeing in Data-Driven Healthcare Workplace: Identity, Agency, and Social Justice},
  author={Sun, Yuling and Ma, Xiaojuan and Lindtner, Silvia and He, Liang},
  journal={Proceedings of the ACM on Human-Computer Interaction},
  volume={7},
  number={CSCW2},
  pages={1--29},
  year={2023},
  publisher={ACM New York, NY, USA}
}

@inproceedings{kristiansen2018accountability,
  title={Accountability in the blue-collar data-driven workplace},
  author={Kristiansen, Kristian Helbo and Valeur-Meller, Mathias A and Dombrowski, Lynn and Holten Moller, Naja L},
  booktitle={Proceedings of the 2018 CHI conference on human factors in computing systems},
  pages={1--12},
  year={2018}
}

@article{heemsbergen2022introduction,
  title={Introduction to algorithmic antagonisms: Resistance, reconfiguration, and renaissance for computational life},
  author={Heemsbergen, Luke and Trer{\'e}, Emiliano and Pereira, Gabriel},
  journal={Media International Australia},
  volume={183},
  number={1},
  pages={3--15},
  year={2022},
  publisher={SAGE Publications Sage UK: London, England}
}

@article{kellogg2020algorithms,
  title={Algorithms at work: The new contested terrain of control},
  author={Kellogg, Katherine C and Valentine, Melissa A and Christin, Angele},
  journal={Academy of management annals},
  volume={14},
  number={1},
  pages={366--410},
  year={2020},
  publisher={Briarcliff Manor, NY}
}

@article{ettlinger2018algorithmic,
  title={Algorithmic affordances for productive resistance},
  author={Ettlinger, Nancy},
  journal={Big Data \& Society},
  volume={5},
  number={1},
  pages={2053951718771399},
  year={2018},
  publisher={SAGE Publications Sage UK: London, England}
}

@article{lake2009relational,
  title={Relational authority and legitimacy in international relations},
  author={Lake, David A},
  journal={American Behavioral Scientist},
  volume={53},
  number={3},
  pages={331--353},
  year={2009},
  publisher={SAGE Publications Sage CA: Los Angeles, CA}
}

@incollection{tyler1992relational,
  title={A relational model of authority in groups},
  author={Tyler, Tom R and Lind, E Allan},
  booktitle={Advances in experimental social psychology},
  volume={25},
  pages={115--191},
  year={1992},
  publisher={Elsevier}
}

@article{sadeghian2024soul,
  title={The soul of work: Evaluation of job meaningfulness and accountability in human-AI collaboration},
  author={Sadeghian, Shadan and Uhde, Alarith and Hassenzahl, Marc},
  journal={Proceedings of the ACM on Human-Computer Interaction},
  volume={8},
  number={CSCW1},
  pages={1--26},
  year={2024},
  publisher={ACM New York, NY, USA}
}

@article{duan2024understanding,
  title={Understanding the evolvement of trust over time within Human-AI teams},
  author={Duan, Wen and Zhou, Shiwen and Scalia, Matthew J and Yin, Xiaoyun and Weng, Nan and Zhang, Ruihao and Freeman, Guo and McNeese, Nathan and Gorman, Jamie and Tolston, Michael},
  journal={Proceedings of the ACM on Human-Computer Interaction},
  volume={8},
  number={CSCW2},
  pages={1--31},
  year={2024},
  publisher={ACM New York, NY, USA}
}

@article{flathmann2024empirically,
  title={Empirically understanding the potential impacts and process of social influence in human-AI teams},
  author={Flathmann, Christopher and Duan, Wen and Mcneese, Nathan J and Hauptman, Allyson and Zhang, Rui},
  journal={Proceedings of the ACM on Human-Computer Interaction},
  volume={8},
  number={CSCW1},
  pages={1--32},
  year={2024},
  publisher={ACM New York, NY, USA}
}

@article{zhang2021ideal,
  title={" An ideal human" expectations of AI teammates in human-AI teaming},
  author={Zhang, Rui and McNeese, Nathan J and Freeman, Guo and Musick, Geoff},
  journal={Proceedings of the ACM on Human-Computer Interaction},
  volume={4},
  number={CSCW3},
  pages={1--25},
  year={2021},
  publisher={ACM New York, NY, USA}
}

@inproceedings{10.1145/3544548.3580959,
    author = {Capel, Tara and Brereton, Margot},
    title = {What is Human-Centered about Human-Centered AI? A Map of the Research Landscape},
    year = {2023},
    isbn = {9781450394215},
    publisher = {Association for Computing Machinery},
    address = {New York, NY, USA},
    url = {https://doi.org/10.1145/3544548.3580959},
    doi = {10.1145/3544548.3580959},
    booktitle = {Proceedings of the 2023 CHI Conference on Human Factors in Computing Systems},
    articleno = {359},
    numpages = {23},
    location = {Hamburg, Germany},
    series = {CHI '23}
}

@inproceedings{lyckvi2018role,
  title={The role of design fiction in participatory design processes},
  author={Lyckvi, Sus and Roto, Virpi and Buie, Elizabeth and Wu, Yiying},
  booktitle={Proceedings of the 10th Nordic conference on human-computer interaction},
  pages={976--979},
  year={2018}
}

@inproceedings{muller2020understanding,
  title={Understanding the past, present, and future of design fictions},
  author={Muller, Michael and Bardzell, Jeffrey and Cheon, EunJeong and Su, Norman Makoto and Baumer, Eric PS and Fiesler, Casey and Light, Ann and Blythe, Mark},
  booktitle={Extended Abstracts of the 2020 CHI Conference on Human Factors in Computing Systems},
  pages={1--8},
  year={2020}
}

@article{sluss2007relational,
  title={Relational identity and identification: Defining ourselves through work relationships},
  author={Sluss, David M and Ashforth, Blake E},
  journal={Academy of management review},
  volume={32},
  number={1},
  pages={9--32},
  year={2007},
  publisher={Academy of Management Briarcliff Manor, NY 10510}
}

@article{raisch2021artificial,
  title={Artificial intelligence and management: The automation--augmentation paradox},
  author={Raisch, Sebastian and Krakowski, Sebastian},
  journal={Academy of management review},
  volume={46},
  number={1},
  pages={192--210},
  year={2021},
  publisher={Academy of Management Briarcliff Manor, NY}
}

@book{west2018future,
  title={The future of work: Robots, AI, and automation},
  author={West, Darrell M},
  year={2018},
  publisher={Brookings Institution Press}
}

@article{shao2025future,
  title={Future of Work with AI Agents: Auditing Automation and Augmentation Potential across the US Workforce},
  author={Shao, Yijia and Zope, Humishka and Jiang, Yucheng and Pei, Jiaxin and Nguyen, David and Brynjolfsson, Erik and Yang, Diyi},
  journal={arXiv preprint arXiv:2506.06576},
  year={2025}
}

@article{zaidi2025will,
  title={How will artificial intelligence (AI) evolve organizational leadership? Understanding the perspectives of technopreneurs},
  author={Zaidi, Syed Yasir Abbas and Aslam, Muhammad Faisal and Mahmood, Faisal and Ahmad, Bilal and Raza, Sadia Bint},
  journal={Global Business and Organizational Excellence},
  volume={44},
  number={3},
  pages={66--83},
  year={2025},
  publisher={Wiley Online Library}
}

@article{prassl2019if,
  title={What if your boss was an algorithm? Economic Incentives, Legal Challenges, and the Rise of Artificial Intelligence at Work},
  author={Prassl, Jeremias},
  journal={Comparative Labor Law and Policy Journal},
  volume={41},
  number={1},
  year={2019},
  publisher={University of Illinois College of Law}
}

@article{lee2019procedural,
  title={Procedural justice in algorithmic fairness: Leveraging transparency and outcome control for fair algorithmic mediation},
  author={Lee, Min Kyung and Jain, Anuraag and Cha, Hea Jin and Ojha, Shashank and Kusbit, Daniel},
  journal={Proceedings of the ACM on human-computer interaction},
  volume={3},
  number={CSCW},
  pages={1--26},
  year={2019},
  publisher={ACM New York, NY, USA}
}

@article{lee2016algorithmic,
  title={Algorithmic bosses, robotic colleagues: toward human-centered algorithmic workplaces},
  author={Lee, Min Kyung},
  journal={XRDS: Crossroads, The ACM Magazine for Students},
  volume={23},
  number={2},
  pages={42--47},
  year={2016},
  publisher={ACM New York, NY, USA}
}

@article{jarrahi2020platformic,
  title={Platformic management, boundary resources for gig work, and worker autonomy},
  author={Jarrahi, Mohammad Hossein and Sutherland, Will and Nelson, Sarah Beth and Sawyer, Steve},
  journal={Computer supported cooperative work (CSCW)},
  volume={29},
  number={1},
  pages={153--189},
  year={2020},
  publisher={Springer}
}

@article{mohlmann2021algorithmic,
  title={Algorithmic management of work on online labor platforms: When matching meets control},
  author={M{\"o}hlmann, Mareike and Zalmanson, Lior and Henfridsson, Ola and Gregory, Robert Wayne},
  journal={MIS quarterly},
  volume={45},
  number={4},
  pages={1999--2022},
  year={2021},
  publisher={Management Information Systems Research Center, University of Minnesota}
}

@incollection{lv_advanced_2023,
    address = {Singapore},
    title = {Advanced {Human}–{Computer} {Interaction} {Technology} in {Digital} {Twins}},
    isbn = {978-981-99-4303-6},
    url = {https://doi.org/10.1007/978-981-99-4303-6_7},
    language = {en},
    urldate = {2025-09-09},
    booktitle = {Artificial {Intelligence} in {IoT} and {Cyborgization}},
    publisher = {Springer Nature},
    author = {Lv, Zhihan and Wu, Jingyi and Chen, Dongliang and Gander, Annn Jia},
    editor = {Dhanaraj, Rajesh Kumar and Rawal, Bharat S. and Krishnamoorthi, Sathya and Balusamy, Balamurugan},
    year = {2023},
    doi = {10.1007/978-981-99-4303-6_7},
    pages = {99--123},
}

@inproceedings{gao_effects_2020,
    title = {The {Effects} of {Avatar} {Visibility} on {Behavioral} {Response} with or without {Mirror}-{Visual} {Feedback} in {Virtual} {Environments}},
    url = {https://ieeexplore.ieee.org/abstract/document/9090683},
    doi = {10.1109/VRW50115.2020.00241},
    urldate = {2025-09-09},
    booktitle = {2020 {IEEE} {Conference} on {Virtual} {Reality} and {3D} {User} {Interfaces} {Abstracts} and {Workshops} ({VRW})},
    author = {Gao, BoYu and Lee, Joonwoo and Tu, Huawei and Seong, Wonjun and Kim, HyungSeok},
    month = mar,
    year = {2020},
    pages = {780--781},
}

@article{weidner_systematic_2023,
  author    = {Weidner, Florian and Boettcher, Gerd and Arboleda, Stephanie Arevalo and Diao, Chenyao and Sinani, Luljeta and Kunert, Christian and Gerhardt, Christoph and Broll, Wolfgang and Raake, Alexander},
  title     = {A Systematic Review on the Visualization of Avatars and Agents in AR \& VR displayed using Head-Mounted Displays},
  journal   = {IEEE Transactions on Visualization and Computer Graphics},
  volume    = {29},
  number    = {5},
  pages     = {2596--2606},
  year      = {2023},
  month     = {May},
  doi       = {10.1109/TVCG.2023.3247072},
  url       = {https://ieeexplore.ieee.org/abstract/document/10049669}
}

@inproceedings{vainionpaa_hci_2022,
    address = {New York, NY, USA},
    series = {Academic {Mindtrek} '22},
    title = {{HCI} and {Digital} {Twins} – {A} {Critical} {Look}: {A} {Literature} {Review}},
    isbn = {978-1-4503-9955-5},
    shorttitle = {{HCI} and {Digital} {Twins} – {A} {Critical} {Look}},
    url = {https://dl.acm.org/doi/10.1145/3569219.3569376},
    doi = {10.1145/3569219.3569376},
    urldate = {2025-09-08},
    booktitle = {Proceedings of the 25th {International} {Academic} {Mindtrek} {Conference}},
    publisher = {Association for Computing Machinery},
    author = {Vainionpää, Fanny and Kinnula, Marianne and Kinnula, Atte and Kuutti, Kari and Hosio, Simo},
    month = nov,
    year = {2022},
    pages = {75--88},
}

@article{weiss_cobots_2021,
    title = {Cobots in {Industry} 4.0: {A} {Roadmap} for {Future} {Practice} {Studies} on {Human}–{Robot} {Collaboration}},
    volume = {51},
    issn = {2168-2305},
    shorttitle = {Cobots in {Industry} 4.0},
    url = {https://ieeexplore.ieee.org/abstract/document/9490032},
    doi = {10.1109/THMS.2021.3092684},
    number = {4},
    urldate = {2025-09-09},
    journal = {IEEE Transactions on Human-Machine Systems},
    author = {Weiss, Astrid and Wortmeier, Ann-Kathrin and Kubicek, Bettina},
    month = aug,
    year = {2021},
    pages = {335--345},
}

@inproceedings{lee_now_2011,
    address = {New York, NY, USA},
    series = {{CHI} '11},
    title = {"{Now}, i have a body": uses and social norms for mobile remote presence in the workplace},
    isbn = {978-1-4503-0228-9},
    shorttitle = {"{Now}, i have a body"},
    url = {https://dl.acm.org/doi/10.1145/1978942.1978950},
    doi = {10.1145/1978942.1978950},
    urldate = {2025-09-11},
    booktitle = {Proceedings of the {SIGCHI} {Conference} on {Human} {Factors} in {Computing} {Systems}},
    publisher = {Association for Computing Machinery},
    author = {Lee, Min Kyung and Takayama, Leila},
    month = may,
    year = {2011},
    pages = {33--42},
}

@inproceedings{lee_algorithmic_2017,
	address = {New York, NY, USA},
	series = {{CSCW} '17},
	title = {Algorithmic {Mediation} in {Group} {Decisions}: {Fairness} {Perceptions} of {Algorithmically} {Mediated} vs. {Discussion}-{Based} {Social} {Division}},
	isbn = {978-1-4503-4335-0},
	shorttitle = {Algorithmic {Mediation} in {Group} {Decisions}},
	url = {https://dl.acm.org/doi/10.1145/2998181.2998230},
	doi = {10.1145/2998181.2998230},
	urldate = {2025-08-31},
	booktitle = {Proceedings of the 2017 {ACM} {Conference} on {Computer} {Supported} {Cooperative} {Work} and {Social} {Computing}},
	publisher = {Association for Computing Machinery},
	author = {Lee, Min Kyung and Baykal, Su},
	month = feb,
	year = {2017},
	pages = {1035--1048},
}

@article{lee_procedural_2019,
	title = {Procedural {Justice} in {Algorithmic} {Fairness}: {Leveraging} {Transparency} and {Outcome} {Control} for {Fair} {Algorithmic} {Mediation}},
	volume = {3},
	shorttitle = {Procedural {Justice} in {Algorithmic} {Fairness}},
	url = {https://dl.acm.org/doi/10.1145/3359284},
	doi = {10.1145/3359284},
	number = {CSCW},
	urldate = {2025-09-01},
	journal = {Proc. ACM Hum.-Comput. Interact.},
	author = {Lee, Min Kyung and Jain, Anuraag and Cha, Hea Jin and Ojha, Shashank and Kusbit, Daniel},
	month = nov,
	year = {2019},
	pages = {182:1--182:26},
}

@inproceedings{umbach_non-consensual_2024,
	address = {New York, NY, USA},
	series = {{CHI} '24},
	title = {Non-{Consensual} {Synthetic} {Intimate} {Imagery}: {Prevalence}, {Attitudes}, and {Knowledge} in 10 {Countries}},
	isbn = {979-8-4007-0330-0},
	shorttitle = {Non-{Consensual} {Synthetic} {Intimate} {Imagery}},
	url = {https://dl.acm.org/doi/10.1145/3613904.3642382},
	doi = {10.1145/3613904.3642382},
	urldate = {2025-08-31},
	booktitle = {Proceedings of the 2024 {CHI} {Conference} on {Human} {Factors} in {Computing} {Systems}},
	publisher = {Association for Computing Machinery},
	author = {Umbach, Rebecca and Henry, Nicola and Beard, Gemma Faye and Berryessa, Colleen M.},
	month = may,
	year = {2024},
	pages = {1--20},
}

@inproceedings{wang_one_2025,
	title = {From {One} {Stolen} {Utterance}: {Assessing} the {Risks} of {Voice} {Cloning} in the {AIGC} {Era}},
	shorttitle = {From {One} {Stolen} {Utterance}},
	url = {https://ieeexplore.ieee.org/document/11023497},
	doi = {10.1109/SP61157.2025.00238},
	urldate = {2025-09-01},
	booktitle = {2025 {IEEE} {Symposium} on {Security} and {Privacy} ({SP})},
	author = {Wang, Kun and Chen, Meng and Lu, Li and Feng, Jingwen and Chen, Qianniu and Ba, Zhongjie and Ren, Kui and Chen, Chun},
	month = may,
	year = {2025},
	note = {ISSN: 2375-1207},
	pages = {4663--4681},
}

@inproceedings{kaate_how_2023,
	address = {New York, NY, USA},
	series = {{CHItaly} '23},
	title = {How {Do} {Users} {Perceive} {Deepfake} {Personas}? {Investigating} the {Deepfake} {User} {Perception} and {Its} {Implications} for {Human}-{Computer} {Interaction}},
	isbn = {979-8-4007-0806-0},
	shorttitle = {How {Do} {Users} {Perceive} {Deepfake} {Personas}?},
	url = {https://dl.acm.org/doi/10.1145/3605390.3605397},
	doi = {10.1145/3605390.3605397},
	urldate = {2025-08-31},
	booktitle = {Proceedings of the 15th {Biannual} {Conference} of the {Italian} {SIGCHI} {Chapter}},
	publisher = {Association for Computing Machinery},
	author = {Kaate, Ilkka and Salminen, Joni and Jung, Soon-Gyo and Almerekhi, Hind and Jansen, Bernard J.},
	month = sep,
	year = {2023},
	pages = {1--12},
}

@article{johnson_what_2021,
	title = {What to do about deepfakes},
	volume = {64},
	issn = {0001-0782},
	url = {https://dl.acm.org/doi/10.1145/3447255},
	doi = {10.1145/3447255},
	number = {3},
	urldate = {2025-09-01},
	journal = {Commun. ACM},
	author = {Johnson, Deborah G. and Diakopoulos, Nicholas},
	month = feb,
	year = {2021},
	pages = {33--35},
}

@article{ofem_tool_2025,
	title = {From {Tool} to {Teammate}: {How} {Transparency} and {Autonomy} {Shape} {Trust}, {Power}, and {Accountability} in {Human}–{AI} {Teams}},
	url = {https://doi.org/10.31124/advance.174669531.17222125/v1},
	doi = {10.31124/ADVANCE.174669531.17222125/V1},
	author = {Ofem, Ofem E},
	month = may,
	year = {2025},
}

@inproceedings{dobre_nice_2022,
	address = {New York, NY, USA},
	series = {{CHI} {EA} '22},
	title = {Nice is {Different} than {Good}: {Longitudinal} {Communicative} {Effects} of {Realistic} and {Cartoon} {Avatars} in {Real} {Mixed} {Reality} {Work} {Meetings}},
	isbn = {978-1-4503-9156-6},
	shorttitle = {Nice is {Different} than {Good}},
	url = {https://dl.acm.org/doi/10.1145/3491101.3519628},
	doi = {10.1145/3491101.3519628},
	urldate = {2025-06-19},
	booktitle = {Extended {Abstracts} of the 2022 {CHI} {Conference} on {Human} {Factors} in {Computing} {Systems}},
	publisher = {Association for Computing Machinery},
	author = {Dobre, Georgiana Cristina and Wilczkowiak, Marta and Gillies, Marco and Pan, Xueni and Rintel, Sean},
	year = {2022},
	pages = {1--7},
}

@article{shamekhi_face_2018,
	title = {Face {Value}? {Exploring} the {Effects} of {Embodiment} for a {Group} {Facilitation} {Agent}},
	url = {https://doi.org/10.1145/3173574.3173965},
	doi = {10.1145/3173574.3173965},
	journal = {International Conference on Human Factors in Computing Systems},
	author = {Shamekhi, Ameneh and Liao, Q. V. and Wang, Dakuo and Bellamy, Rachel K. E. and Erickson, Thomas},
	month = apr,
	year = {2018},
}

@article{truby_human_2021,
	title = {Human digital thought clones: the {Holy} {Grail} of artificial intelligence for big data},
	url = {https://doi.org/10.1080/13600834.2020.1850174},
	doi = {10.1080/13600834.2020.1850174},
	journal = {Information \& communications technology law},
	author = {Truby, J. and Brown, R.},
	month = may,
	year = {2021},
}

@article{dennis_ai_2023,
	title = {{AI} {Agents} as {Team} {Members}: {Effects} on {Satisfaction}, {Conflict}, {Trustworthiness}, and {Willingness} to {Work} {With}},
	url = {https://doi.org/10.1080/07421222.2023.2196773},
	doi = {10.1080/07421222.2023.2196773},
	journal = {Journal of Management Information Systems},
	author = {Dennis, A. and Lakhiwal, Akshat and Sachdeva, Agrim},
	month = apr,
	year = {2023},
}

@article{tang_time_2012,
	title = {Time travel proxy: using lightweight video recordings to create asynchronous, interactive meetings},
	url = {https://doi.org/10.1145/2207676.2208725},
	doi = {10.1145/2207676.2208725},
	journal = {CHI},
	author = {Tang, John C. and Marlow, Jennifer and Hoff, Aaron and Roseway, A. and Quinn, K. and Zhao, Chen and Cao, Xiang},
	month = may,
	year = {2012},
}

@article{huang_mirror_2025,
	title = {Mirror to {Companion}: {Exploring} {Roles}, {Values}, and {Risks} of {AI} {Self}-{Clones} through {Story} {Completion}},
	url = {https://doi.org/10.1145/3706598.3713587},
	doi = {10.1145/3706598.3713587},
	journal = {International Conference on Human Factors in Computing Systems},
	author = {Huang, Jessica and Kim, Ig-Jae and Yoon, Dongwook},
	month = apr,
	year = {2025},
}

@inproceedings{cheng_conversational_2025,
	address = {New York, NY, USA},
	series = {{CHI} '25},
	title = {Conversational {Agents} on {Your} {Behalf}: {Opportunities} and {Challenges} of {Shared} {Autonomy} in {Voice} {Communication} for {Multitasking}},
	isbn = {979-8-4007-1394-1},
	shorttitle = {Conversational {Agents} on {Your} {Behalf}},
	url = {https://dl.acm.org/doi/10.1145/3706598.3714017},
	doi = {10.1145/3706598.3714017},
	urldate = {2025-05-10},
	booktitle = {Proceedings of the 2025 {CHI} {Conference} on {Human} {Factors} in {Computing} {Systems}},
	publisher = {Association for Computing Machinery},
	author = {Cheng, Yi Fei and Shirado, Hirokazu and Kasahara, Shunichi},
	year = {2025},
	pages = {1--18},
}

@article{hwang_whose_2024,
	title = {In {Whose} {Voice}?: {Examining} {AI} {Agent} {Representation} of {People} in {Social} {Interaction} through {Generative} {Speech}},
	url = {https://doi.org/10.1145/3643834.3661555},
	doi = {10.1145/3643834.3661555},
	journal = {Designing Interactive Systems Conference},
	author = {Hwang, A. and Siy, John Oliver and Shelby, Renee and Lentz, Alison},
	month = jul,
	year = {2024},
}

@article{lee_speculating_2023,
	title = {Speculating on {Risks} of {AI} {Clones} to {Selfhood} and {Relationships}: {Doppelganger}-phobia, {Identity} {Fragmentation}, and {Living} {Memories}},
	doi = {10.1145/3579524},
	journal = {Proc. ACM Hum. Comput. Interact.},
	author = {Lee, Patrick Yung Kang and Ma, Ning F. and Kim, Ig-Jae and Yoon, Dongwook},
	year = {2023},
}

@article{leong_dittos_2024,
	title = {Dittos: {Personalized}, {Embodied} {Agents} {That} {Participate} in {Meetings} {When} {You} {Are} {Unavailable}},
	volume = {8},
	shorttitle = {Dittos},
	url = {https://dl.acm.org/doi/10.1145/3687033},
	doi = {10.1145/3687033},
	number = {CSCW2},
	urldate = {2025-05-10},
	journal = {Proc. ACM Hum.-Comput. Interact.},
	author = {Leong, Joanne and Tang, John and Cutrell, Edward and Junuzovic, Sasa and Baribault, Gregory Paul and Inkpen, Kori},
	year = {2024},
	pages = {494:1--494:28},
}

@inproceedings{park_generative_2023,
	address = {New York, NY, USA},
	series = {{UIST} '23},
	title = {Generative {Agents}: {Interactive} {Simulacra} of {Human} {Behavior}},
	isbn = {979-8-4007-0132-0},
	shorttitle = {Generative {Agents}},
	url = {https://dl.acm.org/doi/10.1145/3586183.3606763},
	doi = {10.1145/3586183.3606763},
	urldate = {2025-05-10},
	booktitle = {Proceedings of the 36th {Annual} {ACM} {Symposium} on {User} {Interface} {Software} and {Technology}},
	publisher = {Association for Computing Machinery},
	author = {Park, Joon Sung and O'Brien, Joseph and Cai, Carrie Jun and Morris, Meredith Ringel and Liang, Percy and Bernstein, Michael S.},
	year = {2023},
	pages = {1--22},
}

\appendix

\end{document}